\documentclass{ws-procs961x669}            % default, citations in superscript
\begin{document}
\title{ Fermions and bosons in the expanding universe by 
the spin-charge-family theory }

\author{N. S. Manko\v c Bor\v stnik $^*$ 
}

\address{Department of Physics, University of Ljubljana,\\
SI-1000 Ljubljana, Slovenia\\
$^*$E-mail: norma.mankoc@fmf.uni-lj.si\\
www.fmf.uni-lj.si}

\begin{abstract} 
The {\it spin-charge-family} theory, which is a kind of the Kaluza-Klein theories in $d=(13+1)$ --- but  
with the two kinds of the spin connection fields, the gauge fields of the two Clifford algebra objects,
$S^{ab}$ and $\tilde{S}^{ab}$ --- explains all the assumptions of the {\it standard model}: The 
origin of the charges of fermions appearing in one family, the origin and properties of the vector 
gauge fields of these charges, the origin and properties of the families of fermions, the origin of the 
scalar fields observed as the Higgs's scalar and the Yukawa couplings. The theory explains several 
other phenomena like: The origin of the dark matter, of the matter-antimatter asymmetry, the 
 "miraculous" triangle anomaly cancellation in the {\it standard model} and others. 
Since the theory starts at $d=(13+1)$ 
%with a simple action for gauge fields and spinors --- spinors carry only two kinds of spins and interact 
%correspondingly only with the gravity, the vielbeins and the two kinds of the spin connection fields --- 
the question arises how and at which $d$ had our universe started and how it came down to $d=(13+1)$ 
and further to $d=(3+1)$. In this short contribution some answers to these questions are presented.
\end{abstract}


\bodymatter

\section{Introduction}
\label{introduction}

Both standard models, the {\it standard model} of elementary fermion and boson fields and the 
{\it standard cosmological model}, have quite a lot of assumptions, guessed from the properties of
observables. Although in the history physics was and still is (in particular when many degrees of 
freedom are concerned) relying on small theoretical steps, 
confirmed by experiments, there are also a few decisive steps, without which no real further 
progress would be possible. Among such steps there are certainly the {\it general theory of 
relativity} and the {\it standard model}  of elementary fermion and boson fields.  
Both theories enabled much better understanding of our universe and its elementary fields --- 
fermions and bosons.

With more and more accurate experiments is becoming increasingly clear that a new decisive 
step is again needed in the theory of elementary fields as well as in cosmology.

Both theories rely on observed facts built into innovative mathematical models. However, the 
assumptions remain unexplained. 

Among the non understood assumptions of the {\it standard model} of the elementary fields of 
fermions and bosons are: $\,$ i.  The origin of massless family members with their charges related
to spins. $\,$ ii.  The origin of families of fermions. $\,$ iii.  The origin of the massless vector 
gauge fields of the observed charges.  
$\,$ iv. The origin of masses of family members and heavy bosons.  $\,$ v.  The origin of the 
Higgs's scalar and  the Yukawa couplings.  $\,$ vi.  The origin of
matter-antimatter asymmetry.  $\,$ vii.  The origin of the dark matter.  $\;\,$ viii.  The origin of
the electroweak phase transition scale. $\,$ ix. The origin of the colour phase transition scale. And
others.

 Among the non understood assumptions of the {\it cosmological model} are:
$\, a. $ The differences in the origin of the gravity, of the vector gauge fields and the 
(Higgs's) scalars. $\, b. $ The origin of the dark matter, of the matter-antimatter asymmetry of the 
(ordinary) matter. $\, c. $ The appearance of fermions.
$\, d. $ The origin of the inflation of the universe. $\, e. $ While it is known how to quantize 
vector gauge fields, the quantization of gravity is still an open problem. 

The L(arge) H(adron) C(collider) and other accelerators and measuring apparatus produce a huge 
amount of data, the analyzes of which should help to explain the assumptions of both standard models. 
But it looks like so far that the proposed models, relying more or less on small  extensions of the 
standard models, can not offer much help.
The situation in elementary particle physics is reminiscent of the situation in the nuclear physics before
the {\it standard model} of the elementary fields was proposed, opening new insight into physics of 
elementary fermion and boson fields.

The deeper into the history of our universe we are succeeding to look by the observations and 
experiments the more both standard models are becoming entangled, dependent on each other, 
calling for the next step which would offer the explanation for most of the above mentioned 
 non understood assumptions of both standard models. 

The {\it spin-charge-family} theory~\cite{EPJC2017,norma2014MatterAntimatter,IARD2016,%
JMP2015,pikanorma,norma92,norma93,norma94} does answer open questions of the 
{\it standard model} of the elementary fields and also several of cosmology.

\vspace{2mm}

The  {\it spin-charge-family} theory~\cite{EPJC2017,norma2014MatterAntimatter,IARD2016,%
JMP2015} is promising to be the right next step beyond the 
{\it standard model} of elementary fermion and boson fields by offering the explanation for all 
the assumptions of this model. By offering the explanation also for the dark matter and 
matter-antimatter asymmetry the theory makes a new step also in cosmology, in particular since 
it starts at $d\ge5$ with spinors and gravitational fields only --- like the Kaluza-Klein theories (but with 
the two kinds of the spin connection fields, which are the gauge fields of the two kinds of the Clifford 
algebra objects). 
Although there are still several open problems waiting to be solved, common to most of proposed 
theories --- like how do the 
boundary conditions influence the breaking of the starting symmetry of space-time and how to 
quantize gravity in any $d$, while we know how to quantize at least the vector gauge fields in 
$d=(3+1)$ %~\cite{IARD2016,EPJC2017} 
--- the  {\it spin-charge-family} theory is making several predictions (not just stimulated by the 
current experiments what most of predictions do).

The {\it spin-charge-family} theory (Refs.~\citenum{EPJC2017,norma2014MatterAntimatter,IARD2016,%
JMP2015,pikanorma,norma92,norma93,norma94,gmdn07,gn,gn2015,hn02,hn03,discretesym} %norma95,
%gn2013}, %${}^{\rm{and}\, \rm{Refs.}\, \rm{therein}}$
%,NBled2013,NBled2012,pikanorma,portoroz03,norma92,norma93,norma94,norma95,%
%gmdn07,gn,gn2013,gn2015,NPLB,N2014scalarprop} 
and the references therein) starts in $d=(13+1)$:
{\bf i.} with the simple action for spinors, Eq.~(\ref{wholeaction}), which carry two kinds of spins,
{\bf i.a.} the Dirac one described by $\gamma^a$ and manifesting at low energies in $d=(3+1)$  
as spins and all the charges of the observed fermions of one family, Table~\ref{Table so13+1.}, 
%and Fig.~\ref{universe}, 
%
{\bf i.b.} the second one named~\cite{hn03}  (by the author of this paper) 
$\tilde{\gamma}^a$ ($\{ \tilde{\gamma}^a, \gamma^b \}_{+} =0$, 
 Eq.~(\ref{twoclifford})), and manifesting 
at low energies the family quantum numbers of the observed fermions.
{\bf ii.} Spinors interact in $d=(13+1)$ with the gravitational field only,
{\bf ii.a.} the vielbeins and 
{\bf ii.b.} the two kinds of the spin connection 
fields (Refs.~\cite{EPJC2017,IARD2016} and the references therein).  
%${}^{\rm{and}\, \rm{Refs.}\, \rm{therein}}$.
% 
Spin connection fields --- $\omega_{stm}$ ($(s,t)\ge5$, $m=(0,1,2,3,4)$), Eq.~(\ref{wholeaction}) --- 
are the gauge fields of $S^{st}$, Eq.~(\ref{sabtildesab}), and manifest at low energies in $d=(3+1)$ 
as the vector gauge fields (the colour, weak and hyper vector gauge fields are directly or indirectly 
observed vector gauge fields). Spin connections $\omega_{st s'}$ ($(s,t)\ge5, s'=(7,8)$) manifest as 
scalar gauge fields, contributing to the Higgs's scalar and the Yukawa couplings together with the
scalar spin connection gauge fields --- $\tilde{\omega}_{ab s'}$ ($(a,b)=(m, s,t)$, $s'=(7,8)$), 
Eq.~(\ref{wholeaction}) ---  which are the gauge fields of $\tilde{S}^{ab}$, 
Eq.~(\ref{sabtildesab})~\cite{IARD2016,JMP2015,EPJC2017,norma2014MatterAntimatter}.  
Correspondingly these (several) scalar gauge fields determine after the electroweak break masses 
of the families of all the family members and of the heavy bosons
 (Refs.~\citenum{norma2014MatterAntimatter,%
IARD2016,JMP2015,EPJC2017}, and the references therein).
%${}^{\rm{and}\, \rm{Refs.}\, \rm{therein}}$. 

Scalar fields  $\omega_{st s'}$  ($(s,t)\ge5, s'=(9,\cdots,14)$), 
Ref.~\cite{IARD2016}  (and the references therein),
%${}^{\rm{and}\, \rm{Refs.}\, \rm{therein}}$ 
cause transitions from anti-leptons to quarks and anti-quarks into quarks and back. In the presence 
of the condensate of two right handed neutrinos~\cite{norma2014MatterAntimatter,IARD2016} the
matter-antimatter symmetry breaks.

\section{Short presentation of the {\it spin-charge-family} theory}
\label{SCFT}

The {\it spin-charge-family} theory~\cite{JMP2015,norma2014MatterAntimatter,%
%NBled2013,NBled2012,
IARD2016,%
%pikanorma,portoroz03,%norma92,
norma93,norma94,%norma95,
gmdn07,gn%gn2013,gn2015,NPLB,N2014scalarprop
} assumes a simple action, Eq.~(\ref{wholeaction}), in an even dimensional 
space ($d=2n$, $d>5$). $d$ is chosen to be  $(13+1)$, what makes the simple starting action in $d$
to manifest in $d=(3+1)$ in the low energy regime all the observed degrees of freedom, explaining all 
the assumptions of the {\it standard model} as well as other observed phenomena.
Fermions interact with the vielbeins $f^{\alpha}{}_{a}$ and the two kinds of the spin-connection 
fields --- $\omega_{ab \alpha}$ and $\tilde{\omega}_{ab \alpha}$ ---
the  gauge fields of $S^{ab} = \,\frac{i}{4} (\gamma^a\, \gamma^b
- \gamma^b\, \gamma^a)\,$ and $\tilde{S}^{ab} = \,\frac{i}{4} (\tilde{\gamma}^a\, 
\tilde{\gamma}^b - \tilde{\gamma}^b\, \tilde{\gamma}^a)$, respectively, where:
\begin{eqnarray}
{\cal A}\,  &= \int \; d^dx \; E\;\frac{1}{2}\, (\bar{\psi} \, \gamma^a p_{0a} \psi) + h.c. +
%{\mathcal L}_{f} +  
\nonumber\\  
               &  \int \; d^dx \; E\; (\alpha \,R + \tilde{\alpha} \, \tilde{R})\,,%\nonumber\\
               %\end{eqnarray}
%
%\begin{eqnarray}
%{\mathcal L}_f &=& \frac{1}{2}\, (\bar{\psi} \, \gamma^a p_{0a} \psi) + h.c., 
%\nonumber\\
%p_{0a }        &=& f^{\alpha}{}_a p_{0\alpha} + \frac{1}{2E}\,
% \{ p_{\alpha}, E f^{\alpha}{}_a\}_-, 
%\nonumber\\  
%   p_{0\alpha} &=&  p_{\alpha}  - 
%                    \frac{1}{2}  S^{ab} \omega_{ab \alpha} - 
%                    \frac{1}{2}  \tilde{S}^{ab}   \tilde{\omega}_{ab \alpha},                   
%\nonumber\\ 
%R              &=&  \frac{1}{2} \, \{ f^{\alpha [ a} f^{\beta b ]} \;(\omega_{a b \alpha, \beta} 
%- \omega_{c a \alpha}\,\omega^{c}{}_{b \beta}) \} + h.c. \;, 
%\nonumber\\
%\tilde{R}      &=&  \frac{1}{2} \, \{ f^{\alpha [ a} f^{\beta b ]} \
%;(\tilde{\omega}_{a b \alpha,\beta} - 
%\tilde{\omega}_{c a \alpha} \,\tilde{\omega}^{c}{}_{b \beta})\} + h.c.\;, 
\label{wholeaction}
\end{eqnarray}
here $p_{0a } = f^{\alpha}{}_a\, p_{0\alpha} + \frac{1}{2E}\, \{ p_{\alpha},
E f^{\alpha}{}_a\}_- $, 
$ p_{0\alpha} =  p_{\alpha}  - \frac{1}{2} \, S^{ab}\, \omega_{ab \alpha} - 
                    \frac{1}{2} \,  \tilde{S}^{ab} \,  \tilde{\omega}_{ab \alpha} $~\footnote{Whenever 
two indexes are equal the summation over these two is meant.}, 
\begin{eqnarray} 
%\label{RtildeR}                  
R &=&  \frac{1}{2} \, \{ f^{\alpha [ a} f^{\beta b ]} \;(\omega_{a b \alpha, \beta} 
- \omega_{c a \alpha}\,\omega^{c}{}_{b \beta}) \} + h.c., \,\nonumber  \\
\tilde{R}  &=&  \frac{1}{2} \, \{ f^{\alpha [ a} f^{\beta b ]} \;(\tilde{\omega}_{a b \alpha,\beta} - 
\tilde{\omega}_{c a \alpha} \,\tilde{\omega}^{c}{}_{b \beta})\} + h.c.. \nonumber
\end{eqnarray}
 The action introduces two kinds of the Clifford algebra objects, $\gamma^a$ and
 $\tilde{\gamma}^a$,
\begin{eqnarray}
\label{twoclifford}
\{\gamma^a, \gamma^b\}_{+}= 2 \eta^{ab} = 
\{\tilde{\gamma}^a, \tilde{\gamma}^b\}_{+}\,.
\end{eqnarray}
$f^{\alpha}{}_{a}$ are vielbeins inverted to $e^{a}{}_{\alpha}$, Latin letters ($a,b,..$) denote
flat indices, Greek letters ($\alpha,\beta,..$) are Einstein indices,  $(m,n,..)$ and $(\mu,\nu,..)$ 
denote the corresponding indices in ($0,1,2,3$), $(s,t,..)$ and $(\sigma,\tau,..)$  denote the
corresponding indices in $d\ge5$:
\begin{eqnarray}
\label{vielfe}
e^{a}{}_{\alpha}f^{\beta}{}_{a} &=\delta^{\beta}_{\alpha}\,, \quad
e^{a}{}_{\alpha}f^{\alpha}{}_{b}= \delta^{a}_{b}\,,
\end{eqnarray}
$E =\det(e^{a}{}_{\alpha})$~\footnote{This definition of the vielbein and the inverted vielbein is 
general, no specification about the curled space is assumed yet, but is valid also in the low energies 
regions, when the starting symmetry is broken~\cite{EPJC2017}.}.

The action ${\cal A} $ offers the explanation for the origin and all the properties of the observed 
fermions  (of the family members and families), of the observed vector gauge fields, of the Higgs's
scalar and of the Yukawa couplings, explaining the origin of the matter-antimatter asymmetry,  the 
appearance of the dark matter and predicts the new scalars and the new (fourth) family coupled to the
observed three %and a new gauge field 
to be measured at the LHC~(\cite{norma2014MatterAntimatter,IARD2016} and the 
references therein).  

The {\it standard model} groups of spins and charges are the subgroups of the $SO(13,1)$ group 
with the generator of the infinitesimal transformations expressible with $S^{ab}$ --- for spins 
\begin{eqnarray}
\label{so1+3}
\vec{N}_{\pm}(= \vec{N}_{(L,R)}): &= \,\frac{1}{2} (S^{23}\pm i S^{01},S^{31}\pm i S^{02}, 
S^{12}\pm i S^{03} )\,,%\,,\nonumber\\
%\vec{\tilde{N}}_{\pm}(=\vec{\tilde{N}}_{(L,R)}):&=& \,\frac{1}{2} (\tilde{S}^{23}\pm i \tilde{S}^{01},
%\tilde{S}^{31}\pm i \tilde{S}^{02}, \tilde{S}^{12}\pm i \tilde{S}^{03} )\,,
\end{eqnarray}
--- for the weak charge, $SU(2)_{I}$, and the second $SU(2)_{II}$, these two groups are the invariant 
subgroups of $SO(4)$
 \begin{eqnarray}
 \label{so42}
 \vec{\tau}^{1}:&=\frac{1}{2} (S^{58}-  S^{67}, \,S^{57} + S^{68}, \,S^{56}-  S^{78} )\,,
\nonumber\\
 \vec{\tau}^{2}:&= \frac{1}{2} (S^{58}+  S^{67}, \,S^{57} - S^{68}, \,S^{56}+  S^{78} )\,,
%\,\;\nonumber\\
 %\vec{\tilde{\tau}}^{1}:&=&\frac{1}{2} (\tilde{S}^{58}-  \tilde{S}^{67}, \,\tilde{S}^{57} + 
 %\tilde{S}^{68}, \,\tilde{S}^{56}-  \tilde{S}^{78} )\,,\;\;
 %\vec{\tilde{\tau}}^{2}:=\frac{1}{2} (\tilde{S}^{58}+  \tilde{S}^{67}, \,\tilde{S}^{57} - 
 %\tilde{S}^{68}, \,\tilde{S}^{56}+  \tilde{S}^{78} ),\,\,\;\;
 \end{eqnarray}
--- for the colour charge  $SU(3)$ and for the "fermion charge" $U(1)_{II}$, these two groups are
subgroups of $SO(6)$
 \begin{eqnarray}
 \label{so64}
 \vec{\tau}^{3}: &= &\frac{1}{2} \,\{  S^{9\;12} - S^{10\;11} \,,
  S^{9\;11} + S^{10\;12} ,\, S^{9\;10} - S^{11\;12} ,\nonumber\\
 && S^{9\;14} -  S^{10\;13} ,\,  S^{9\;13} + S^{10\;14} \,,
  S^{11\;14} -  S^{12\;13}\,,\nonumber\\
 && S^{11\;13} +  S^{12\;14} ,\, 
 \frac{1}{\sqrt{3}} ( S^{9\;10} + S^{11\;12} - 
 2 S^{13\;14})\}\,,\nonumber\\
 \tau^{4}: &=& -\frac{1}{3}(S^{9\;10} + S^{11\;12} + S^{13\;14})\,,%\;\;\nonumber\\
 %\tilde{\tau}^{4}: = &&-\frac{1}{3}(\tilde{S}^{9\;10} + \tilde{S}^{11\;12} + \tilde{S}^{13\;14})\,,
 \end{eqnarray}
--- while the hyper charge $Y$ is $Y=\tau^{23} + \tau^{4}$. The breaks of the symmetries, manifesting
in Eqs.~(\ref{so1+3}, \ref{so42}, \ref{so64}), are in the {\it spin-charge-family} theory caused by
the condensate and the constant values of the scalar fields carrying the space index $(7,8)$ 
(Refs.~\cite{JMP2015,IARD2016} and the references therein). The space breaks first to $SO(7,1)$
$\times SU(3) \times U(1)_{II}$ and then further to $SO(3,1)\times SU(2)_{I} \times U(1)_{I}$
$\times SU(3)$, what explains the connections between the weak and the hyper 
charges and the handedness of spinors.

The equivalent expressions for the family charges, expressed by $\tilde{S}^{ab}$, follow if in 
Eqs.~(\ref{so1+3} - \ref{so64}) $S^{ab}$ are replaced by $\tilde{S}^{ab}$. 

\subsection{A short inside into the spinor states of the {\it spin-charge-family} theory}
\label{spinorSCFT}

I demonstrate in this subsection on two examples how transparently can properties of spinor and 
anti-spinor states  be read from these states~\cite{hn02,hn03,JMP2015}, when the states are 
expressed with $\frac{d}{2}$ nilpotents and projectors, formed as odd and even objects of 
$\gamma^a$'s (Eq.~(\ref{signature})) and chosen to be the eigenstates of the Cartan 
subalgebra (Eq.~(\ref{choicecartan})) of the algebra of the two groups, as in 
Table~\ref{Table so13+1.}.

Recognizing that the two Clifford algebra objects $(S^{ab},S^{cd}$), 
or ($\tilde{S}^{ab},\tilde{S}^{cd}$), fulfilling the algebra,
\begin{eqnarray}
\label{sabtildesab}
\{S^{ab},S^{cd}\}_{-} &=&  
 i(\eta^{ad} S^{bc} + \eta^{bc} S^{ad} - \eta^{ac} S^{bd} - \eta^{bd} S^{ac})\,,\nonumber\\
\{\tilde{S}^{ab},\tilde{S}^{cd}\}_{-} &=& i(\eta^{ad} \tilde{S}^{bc} + \eta^{bc} \tilde{S}^{ad} 
- \eta^{ac} \tilde{S}^{bd} - \eta^{bd} \tilde{S}^{ac})\,,\nonumber\\ 
 \{S^{ab}, \tilde{S}^{cd}\}_{-}&=& 0\,, 
\end{eqnarray} 
commute,  if %$\{S^{ab},S^{cd}\}_{-}$ and  $\{\tilde{S}^{ab},\tilde{S}^{cd}\}_{-}$ 
all the indexes ($a,b,c,d$) are different, the Cartan subalgebra is in 
$d=2n$  selected as follows 
\begin{eqnarray}
S^{03}, S^{12}, S^{56}, \cdots, S^{d-1\; d}, \quad {\rm if } \quad d &= 2n\ge 4\,,
%\nonumber\\
%S^{03}, S^{12}, \cdots, S^{d-2 \;d-1}, \quad {\rm if } \quad d &=& (2n +1) >4\,,
\nonumber\\
\tilde{S}^{03}, \tilde{S}^{12}, \tilde{S}^{56}, \cdots, \tilde{S}^{d-1\; d}\,, 
\quad {\rm if } \quad d &= 2n\ge 4\,.
%\nonumber\\
%\tilde{S}^{03}, \tilde{S}^{12}, \cdots, \tilde{S}^{d-2 \;d-1}, 
%\quad {\rm if } \quad d &= (2n +1) >4\,.
\label{choicecartan}
\end{eqnarray}
Let us define as well one of the Casimirs of the Lorentz group ---  
the  handedness $\Gamma$ ($\{\Gamma, S^{ab}\}_- =0$) in $d=2n$~\footnote{The reader
can find the definition of handedness for $d$ odd in Refs.~\cite{hn02,IARD2016} and the references 
therein.}
\begin{eqnarray}
\Gamma^{(d)} :&=(i)^{d/2}\,\prod_a  (\sqrt{\eta^{aa}}
 \gamma^a), \quad {\rm if } \quad d = 2n\,, 
%\nonumber\\
%\Gamma^{(d)} :&= (i)^{(d-1)/2}\; \prod_a \quad (\sqrt{\eta^{aa}} \gamma^a), \quad {\rm if } \quad d = 2n +1\,.
\label{hand}
\end{eqnarray}
which can be written also as $\Gamma^{(d)} = i^{d-1} \cdot 2^{\frac{d}{2}}$ $S^{03}\cdot S^{12}
\cdots S^{(d-1) \,d}$.
The product of $\gamma^a$'s must be taken in the ascending order with 
respect to the index $a$: $\gamma^0 \gamma^1\cdots \gamma^d$. 
It follows from the Hermiticity properties of $\gamma^a $ %Eq.(\ref{cliffher})
for any choice of the signature $\eta^{aa}$ that $\Gamma^{(d)\dagger}= \Gamma^{(d)},\;
(\Gamma^{(d)})^2 = I$. One proceeds equivalently for $\tilde{\Gamma}^{(d)} $, substituting 
$\gamma^a$'s by $\tilde{\gamma}^a$'s.
We also find that for $d$ even the handedness  anticommutes with the Clifford algebra objects 
$\gamma^a$ ($\{\gamma^a, \Gamma \}_+ = 0$).
% while for $d$ odd it commutes with $\gamma^a$ ($\{\gamma^a, \Gamma \}_- = 0$). 

Spinor states can be, as in Table~\ref{Table so13+1.}, represented as products of nilpotents and 
projectors defined by $\gamma^a$'s
\begin{eqnarray}
\stackrel{ab}{(k)}:&=
\frac{1}{2}(\gamma^a + \frac{\eta^{aa}}{ik} \gamma^b)\,,\quad \quad
\stackrel{ab}{[k]}:=
\frac{1}{2}(1+ \frac{i}{k} \gamma^a \gamma^b)\,,%\nonumber\\
%\stackrel{+}{\circ}:&=& \frac{1}{2} (1+\Gamma)\,,\quad \quad
%\stackrel{-}{\bullet}:= \frac{1}{2}(1-\Gamma),
\label{signature}
\end{eqnarray}
where $k^2 = \eta^{aa} \eta^{bb}$. 

It is easy to check that the nilpotent $\stackrel{ab}{(k)}$ and the projector $\stackrel{ab}{[k]}$ 
are "eigenstates" of $S^{ab}$ and $\tilde{S}^{ab}$ 
\begin{eqnarray}
        &S^{ab}\, \stackrel{ab}{(k)}= \frac{1}{2}\,k\, \stackrel{ab}{(k)}\,,\quad \quad 
        S^{ab}\, \stackrel{ab}{[k]}= \;\;\frac{1}{2}\,k \,\stackrel{ab}{[k]}\,,\nonumber\\
&\tilde{S}^{ab}\, \stackrel{ab}{(k)}= \frac{1}{2}\,k \,\stackrel{ab}{(k)}\,,\quad \quad 
\tilde{S}^{ab}\, \stackrel{ab}{[k]}=-\frac{1}{2}\,k \,\stackrel{ab}{[k]}\,,
\label{grapheigen}
\end{eqnarray}
where in Eq.~(\ref{grapheigen}) the vacuum state $|\psi_0\rangle$ is meant to stay on the
right hand sides of projectors 
and nilpotents. This means that one gets when multiplying nilpotents $\stackrel{ab}{(k)}$ and
 projectors $\stackrel{ab}{[k]}$ by $S^{ab}$  the same objects back multiplied by the 
constant $\frac{1}{2}k$, while $\tilde{S}^{ab}$ multiply $\stackrel{ab}{(k)}$ by $k$ 
and $\stackrel{ab}{[k]}$ by $(-k)$ rather than $k$.

One can namely see,  taking into account Eq.~(\ref{twoclifford}), that
\begin{eqnarray}
%\label{snmb:graphgammatilgegammaaction}
&\gamma^a \stackrel{ab}{(k)}= \eta^{aa}\stackrel{ab}{[-k]},\; 
\gamma^b \stackrel{ab}{(k)}= -ik \stackrel{ab}{[-k]}, \; 
\gamma^a \stackrel{ab}{[k]}= \stackrel{ab}{(-k)},\; 
\gamma^b \stackrel{ab}{[k]}= -ik \eta^{aa} \stackrel{ab}{(-k)}\,,\nonumber\\
&\tilde{\gamma^a} \stackrel{ab}{(k)} = - i\eta^{aa}\stackrel{ab}{[k]},\;
\tilde{\gamma^b} \stackrel{ab}{(k)} =  - k \stackrel{ab}{[k]}, \;
\tilde{\gamma^a} \stackrel{ab}{[k]} =  \;\;i\stackrel{ab}{(k)},\; 
\tilde{\gamma^b} \stackrel{ab}{[k]} =  -k \eta^{aa} \stackrel{ab}{(k)}\,. 
\label{snmb:gammatildegamma}
\end{eqnarray}
One recognizes %~(Eq.~\ref{graphgammaaction},\ref{gammatilde}) 
 also that $\gamma^a$ 
transform  $\stackrel{ab}{(k)}$ into  $\stackrel{ab}{[-k]}$, never to $\stackrel{ab}{[k]}$, 
while $\tilde{\gamma}^a$ transform  $\stackrel{ab}{(k)}$ into $\stackrel{ab}{[k]}$, never to 
$\stackrel{ab}{[-k]}$. 
%%%%%%%%%%%%%%%%%%%%%%%12.04.

In Table~\ref{Table so13+1.}~\cite{norma2014MatterAntimatter,pikanorma,JMP2015}
the left handed ($\Gamma^{(13,1)} = -1$, Eq.~(\ref{hand})) massless multiplet of one 
family (Table~\ref{Table III.}) of spinors --- the members of the fundamental representation of the 
$SO(13,1)$ group ---  is presented as products of nilpotents and projectors, Eq.~(\ref{signature}). 
All these states are eigenstates  of the Cartan sub-algebra (Eq.~(\ref{choicecartan})). 
Table~\ref{Table so13+1.} manifests the subgroup $SO(7,1)$  of the colour charged quarks and 
anti-quarks and the colourless leptons and anti-leptons~\cite{hn02,hn03}. The multiplet contains 
the left handed  ($\Gamma^{(3,1)}=-1$) weak ($SU(2)_{I}$) charged  
($\tau^{13}=\pm \frac{1}{2}$, Eq.~(\ref{so42})), 
%$\vec{\tau}^{1}= \frac{1}{2} (S^{58}- S^{67}, S^{57}+ S^{68}, S^{56}- S^{78})$) 
and $SU(2)_{II}$ chargeless ($\tau^{23}=0$, Eq.~(\ref{so42}))
%$\vec{\tau}^{2}= \frac{1}{2} (S^{58}+ S^{67}, S^{57}- S^{68}, S^{56}+ S^{78})$)
quarks and leptons and the right handed  ($\Gamma^{(3,1)}=1$)  weak  ($SU(2)_{I}$) 
chargeless and $SU(2)_{II}$
charged ($\tau^{23}=\pm \frac{1}{2}$) quarks and leptons, both with the spin $ S^{12}$  up and 
down ($\pm \frac{1}{2}$, respectively). 
Quarks and leptons (and separately anti-quarks and anti-leptons) have the same $SO(7,1)$ 
part. They distinguish only in the $SU(3) \times U(1)$ part: Quarks are triplets 
of three colours  ($c^i$ $= (\tau^{33}, \tau^{38})$ $ = [(\frac{1}{2},\frac{1}{2\sqrt{3}}), 
(-\frac{1}{2},\frac{1}{2\sqrt{3}}), (0,-\frac{1}{\sqrt{3}}) $], Eq.~(\ref{so64}))
%$\vec{\tau}^{3}= \frac{1}{2}(S^{9\,12}- S^{10\,11},S^{9\,11}+ S^{10\,12},S^{9\,10}- 
%S^{11\,12},$ $S^{9\,14}- S^{10\,13},S^{9\,13}+ S^{10\,14},S^{11\,14}- S^{12\,13},$
%$S^{11\,13}+ S^{12\,14},\frac{1}{\sqrt{3}}(S^{9\,10}+ S^{11\,12} - 2S^{13\,14})$), 
carrying  the "fermion charge" ($\tau^{4}=\frac{1}{6}$, Eq.~(\ref{so64})).
%$=-\frac{1}{3}(S^{9\,10}+ S^{11\,12}+ S^{13\,14})$).
The colourless leptons carry the "fermion charge" ($\tau^{4}=-\frac{1}{2}$). 

The same multiplet contains also the left handed weak ($SU(2)_{I}$) chargeless and $SU(2)_{II}$ 
charged anti-quarks and anti-leptons and the right handed weak ($SU(2)_{I}$) charged and 
$SU(2)_{II}$ chargeless anti-quarks and anti-leptons. 
%Anti-quarks distinguish from anti-leptons again only in the $SU(3) \times U(1)$ part:
 Anti-quarks are anti-triplets, % ($\bar{c}^i$ $= (\tau^{33}, \tau^{38})$), 
 carrying  the "fermion charge" ($\tau^{4}=-\frac{1}{6}$). 
%$=-\frac{1}{3}(S^{9\,10}+ S^{11\,12}+ S^{13\,14})$)
The anti-colourless anti-leptons carry the "fermion charge" ($\tau^{4}=\frac{1}{2}$). 
$ S^{12}$ defines the ordinary spin %(which can also be read directly from the basic vector, both
%vectors  with both spins, 
$\pm \frac{1}{2}$.  $Y=(\tau^{23} + \tau^{4})$ is the hyper charge, the electromagnetic charge 
is $Q=(\tau^{13} + Y$).
The vacuum state, 
%$|vac>_{fam}$, 
on which the nilpotents and projectors operate, is not shown. 

All these properties of states can be read directly from the table. {\it Example 1. and 2.}
demonstrate how this can be done.

The states of opposite charges (anti-particle states) are reachable  from the particle states 
(besides by
$S^{ab}$) also by the application of the discrete symmetry operator 
${\cal C}_{{\cal N}}$ ${\cal P}_{{\cal N}}$, presented in Refs.~\cite{discretesym} and in the 
footnote of this subsection. %~\ref{spinorSCFT}.
%

%%%%%%%%%%%%%%%%%%%%

%\begin{tiny} 
\begin{table}
\begin{tiny}
\tbl{The left handed ($\Gamma^{(13,1)} = -1$, 
% ($ = \Gamma^{(7,1)} \times \Gamma^{(6)}$),  
Eq.~(\ref{hand})) 
multiplet of spinors --- the members of (one family of) the fundamental representation of the 
$SO(13,1)$ group %manifesting the subgroup $SO(7,1)$ 
of the colour charged quarks and anti-quarks and the colourless leptons and anti-leptons, with the
charges, spin and handedness manifesting in the low energy regime ---
 is presented in the massless basis using the
 technique~\cite{norma2014MatterAntimatter,pikanorma,JMP2015}, explained in the 
text and in {\it Examples 1.,2.}.} 
%\end{tiny}
%\begin{tiny}
%{\begin{tabular}{|r|c ||c||c|c||c|c||c|c|c||r|r|} 
     {\begin{tabular}{r c |c  |r  r | r  r | r r r | r r }
%\begin{tiny}
\toprule
i&$$&$|^a\psi_i>$, $\,\Gamma^{(7,1)} = (-1)\,1\,, \,\Gamma^{(6)} = (1)\,-1$ &$\Gamma^{(3,1)}$&$ S^{12}$&
$\tau^{13}$&$\tau^{23}$&$\tau^{33}$&$\tau^{38}$&$\tau^{4}$&$Y$&$Q$\\
%\hline
&& ${\rm (Anti)octet}$&&&&&&&&& \\
%,\,\Gamma^{(7,1)} = (-1)\,1\,, \,\Gamma^{(6)} = (1)\,-1
&& ${\rm of \;(anti) quarks \;and \;(anti)leptons}$&&&&&&&&&\\
%\hline
\hline
1&$ u_{R}^{c1}$&$ \stackrel{03}{(+i)}\,\stackrel{12}{(+)}|
\stackrel{56}{(+)}\,\stackrel{78}{(+)}
||\stackrel{9 \;10}{(+)}\;\;\stackrel{11\;12}{(-)}\;\;\stackrel{13\;14}{(-)} $ &1&$\frac{1}{2}$&0&
$\frac{1}{2}$&$\frac{1}{2}$&$\frac{1}{2\,\sqrt{3}}$&$\frac{1}{6}$&$\frac{2}{3}$&$\frac{2}{3}$\\
%\hline 
2&$u_{R}^{c1}$&$\stackrel{03}{[-i]}\,\stackrel{12}{[-]}|\stackrel{56}{(+)}\,\stackrel{78}{(+)}
||\stackrel{9 \;10}{(+)}\;\;\stackrel{11\;12}{(-)}\;\;\stackrel{13\;14}{(-)}$&1&$-\frac{1}{2}$&0&
$\frac{1}{2}$&$\frac{1}{2}$&$\frac{1}{2\,\sqrt{3}}$&$\frac{1}{6}$&$\frac{2}{3}$&$\frac{2}{3}$\\
%\hline
3&$d_{R}^{c1}$&$\stackrel{03}{(+i)}\,\stackrel{12}{(+)}|\stackrel{56}{[-]}\,\stackrel{78}{[-]}
||\stackrel{9 \;10}{(+)}\;\;\stackrel{11\;12}{(-)}\;\;\stackrel{13\;14}{(-)}$&1&$\frac{1}{2}$&0&
$-\frac{1}{2}$&$\frac{1}{2}$&$\frac{1}{2\,\sqrt{3}}$&$\frac{1}{6}$&$-\frac{1}{3}$&$-\frac{1}{3}$\\
%\hline 
4&$ d_{R}^{c1} $&$\stackrel{03}{[-i]}\,\stackrel{12}{[-]}|
\stackrel{56}{[-]}\,\stackrel{78}{[-]}
||\stackrel{9 \;10}{(+)}\;\;\stackrel{11\;12}{(-)}\;\;\stackrel{13\;14}{(-)} $&1&$-\frac{1}{2}$&0&
$-\frac{1}{2}$&$\frac{1}{2}$&$\frac{1}{2\,\sqrt{3}}$&$\frac{1}{6}$&$-\frac{1}{3}$&$-\frac{1}{3}$\\
%\hline
5&$d_{L}^{c1}$&$\stackrel{03}{[-i]}\,\stackrel{12}{(+)}|\stackrel{56}{[-]}\,\stackrel{78}{(+)}
||\stackrel{9 \;10}{(+)}\;\;\stackrel{11\;12}{(-)}\;\;\stackrel{13\;14}{(-)}$&-1&$\frac{1}{2}$&
$-\frac{1}{2}$&0&$\frac{1}{2}$&$\frac{1}{2\,\sqrt{3}}$&$\frac{1}{6}$&$\frac{1}{6}$&$-\frac{1}{3}$\\
%\hline
6&$d_{L}^{c1} $&$\stackrel{03}{(+i)}\,\stackrel{12}{[-]}|\stackrel{56}{[-]}\,\stackrel{78}{(+)}
||\stackrel{9 \;10}{(+)}\;\;\stackrel{11\;12}{(-)}\;\;\stackrel{13\;14}{(-)} $&-1&$-\frac{1}{2}$&
$-\frac{1}{2}$&0&$\frac{1}{2}$&$\frac{1}{2\,\sqrt{3}}$&$\frac{1}{6}$&$\frac{1}{6}$&$-\frac{1}{3}$\\
%\hline
7&$ u_{L}^{c1}$&$\stackrel{03}{[-i]}\,\stackrel{12}{(+)}|\stackrel{56}{(+)}\,\stackrel{78}{[-]}
||\stackrel{9 \;10}{(+)}\;\;\stackrel{11\;12}{(-)}\;\;\stackrel{13\;14}{(-)}$ &-1&$\frac{1}{2}$&
$\frac{1}{2}$&0 &$\frac{1}{2}$&$\frac{1}{2\,\sqrt{3}}$&$\frac{1}{6}$&$\frac{1}{6}$&$\frac{2}{3}$\\
%\hline
8&$u_{L}^{c1}$&$\stackrel{03}{(+i)}\,\stackrel{12}{[-]}|\stackrel{56}{(+)}\,\stackrel{78}{[-]}
||\stackrel{9 \;10}{(+)}\;\;\stackrel{11\;12}{(-)}\;\;\stackrel{13\;14}{(-)}$&-1&$-\frac{1}{2}$&
$\frac{1}{2}$&0&$\frac{1}{2}$&$\frac{1}{2\,\sqrt{3}}$&$\frac{1}{6}$&$\frac{1}{6}$&$\frac{2}{3}$\\
%\hline
\hline
%\shrinkheight{0.25\textheight}
9&$ u_{R}^{c2}$&$ \stackrel{03}{(+i)}\,\stackrel{12}{(+)}|
\stackrel{56}{(+)}\,\stackrel{78}{(+)}
||\stackrel{9 \;10}{[-]}\;\;\stackrel{11\;12}{[+]}\;\;\stackrel{13\;14}{(-)} $ &1&$\frac{1}{2}$&0&
$\frac{1}{2}$&$-\frac{1}{2}$&$\frac{1}{2\,\sqrt{3}}$&$\frac{1}{6}$&$\frac{2}{3}$&$\frac{2}{3}$\\
%\hline 
10&$u_{R}^{c2}$&$\stackrel{03}{[-i]}\,\stackrel{12}{[-]}|\stackrel{56}{(+)}\,\stackrel{78}{(+)}
||\stackrel{9 \;10}{[-]}\;\;\stackrel{11\;12}{[+]}\;\;\stackrel{13\;14}{(-)}$&1&$-\frac{1}{2}$&0&
$\frac{1}{2}$&$-\frac{1}{2}$&$\frac{1}{2\,\sqrt{3}}$&$\frac{1}{6}$&$\frac{2}{3}$&$\frac{2}{3}$\\
%\hline
%$\cdots$&&&&&&&&&&&\\
%\hline 
11&$d_{R}^{c2}$&$\stackrel{03}{(+i)}\,\stackrel{12}{(+)}|\stackrel{56}{[-]}\,\stackrel{78}{[-]}
||\stackrel{9 \;10}{[-]}\;\;\stackrel{11\;12}{[+]}\;\;\stackrel{13\;14}{(-)}$
&1&$\frac{1}{2}$&0&
$-\frac{1}{2}$&$-\frac{1}{2}$&$\frac{1}{2\,\sqrt{3}}$&$\frac{1}{6}$&$-\frac{1}{3}$&$-\frac{1}{3}$\\
%\hline 
12&$ d_{R}^{c2} $&$\stackrel{03}{[-i]}\,\stackrel{12}{[-]}|
\stackrel{56}{[-]}\,\stackrel{78}{[-]}
||\stackrel{9 \;10}{[-]}\;\;\stackrel{11\;12}{[+]}\;\;\stackrel{13\;14}{(-)} $
&1&$-\frac{1}{2}$&0&
$-\frac{1}{2}$&$-\frac{1}{2}$&$\frac{1}{2\,\sqrt{3}}$&$\frac{1}{6}$&$-\frac{1}{3}$&$-\frac{1}{3}$\\
%\hline
13&$d_{L}^{c2}$&$\stackrel{03}{[-i]}\,\stackrel{12}{(+)}|\stackrel{56}{[-]}\,\stackrel{78}{(+)}
||\stackrel{9 \;10}{[-]}\;\;\stackrel{11\;12}{[+]}\;\;\stackrel{13\;14}{(-)}$
&-1&$\frac{1}{2}$&
$-\frac{1}{2}$&0&$-\frac{1}{2}$&$\frac{1}{2\,\sqrt{3}}$&$\frac{1}{6}$&$\frac{1}{6}$&$-\frac{1}{3}$\\
%\hline
14&$d_{L}^{c2} $&$\stackrel{03}{(+i)}\,\stackrel{12}{[-]}|\stackrel{56}{[-]}\,\stackrel{78}{(+)}
||\stackrel{9 \;10}{[-]}\;\;\stackrel{11\;12}{[+]}\;\;\stackrel{13\;14}{(-)} $&-1&$-\frac{1}{2}$&
$-\frac{1}{2}$&0&$-\frac{1}{2}$&$\frac{1}{2\,\sqrt{3}}$&$\frac{1}{6}$&$\frac{1}{6}$&$-\frac{1}{3}$\\
%\hline
15&$ u_{L}^{c2}$&$\stackrel{03}{[-i]}\,\stackrel{12}{(+)}|\stackrel{56}{(+)}\,\stackrel{78}{[-]}
||\stackrel{9 \;10}{[-]}\;\;\stackrel{11\;12}{[+]}\;\;\stackrel{13\;14}{(-)}$ &-1&$\frac{1}{2}$&
$\frac{1}{2}$&0 &$-\frac{1}{2}$&$\frac{1}{2\,\sqrt{3}}$&$\frac{1}{6}$&$\frac{1}{6}$&$\frac{2}{3}$\\
%\hline
16&$u_{L}^{c2}$&$\stackrel{03}{(+i)}\,\stackrel{12}{[-]}|\stackrel{56}{(+)}\,\stackrel{78}{[-]}
||\stackrel{9 \;10}{[-]}\;\;\stackrel{11\;12}{[+]}\;\;\stackrel{13\;14}{(-)}$&-1&$-\frac{1}{2}$&
$\frac{1}{2}$&0&$-\frac{1}{2}$&$\frac{1}{2\,\sqrt{3}}$&$\frac{1}{6}$&$\frac{1}{6}$&$\frac{2}{3}$\\
%\hline
\hline
17&$ u_{R}^{c3}$&$ \stackrel{03}{(+i)}\,\stackrel{12}{(+)}|
\stackrel{56}{(+)}\,\stackrel{78}{(+)}
||\stackrel{9 \;10}{[-]}\;\;\stackrel{11\;12}{(-)}\;\;\stackrel{13\;14}{[+]} $ &1&$\frac{1}{2}$&0&
$\frac{1}{2}$&$0$&$-\frac{1}{\sqrt{3}}$&$\frac{1}{6}$&$\frac{2}{3}$&$\frac{2}{3}$\\
%\hline 
18&$u_{R}^{c3}$&$\stackrel{03}{[-i]}\,\stackrel{12}{[-]}|\stackrel{56}{(+)}\,\stackrel{78}{(+)}
||\stackrel{9 \;10}{[-]}\;\;\stackrel{11\;12}{(-)}\;\;\stackrel{13\;14}{[+]}$&1&$-\frac{1}{2}$&0&
$\frac{1}{2}$&$0$&$-\frac{1}{\sqrt{3}}$&$\frac{1}{6}$&$\frac{2}{3}$&$\frac{2}{3}$\\
%\hline
%$\cdots$&&&&&&&&&&\\
19&$d_{R}^{c3}$&$\stackrel{03}{(+i)}\,\stackrel{12}{(+)}|\stackrel{56}{[-]}\,\stackrel{78}{[-]}
||\stackrel{9 \;10}{[-]}\;\;\stackrel{11\;12}{(-)}\;\;\stackrel{13\;14}{[+]}$&1&$\frac{1}{2}$&0&
$-\frac{1}{2}$&$0$&$-\frac{1}{\sqrt{3}}$&$\frac{1}{6}$&$-\frac{1}{3}$&$-\frac{1}{3}$\\
%\hline 
20&$ d_{R}^{c3} $&$\stackrel{03}{[-i]}\,\stackrel{12}{[-]}|
\stackrel{56}{[-]}\,\stackrel{78}{[-]}
||\stackrel{9 \;10}{[-]}\;\;\stackrel{11\;12}{(-)}\;\;\stackrel{13\;14}{[+]} $&1&$-\frac{1}{2}$&0&
$-\frac{1}{2}$&$0$&$-\frac{1}{\sqrt{3}}$&$\frac{1}{6}$&$-\frac{1}{3}$&$-\frac{1}{3}$\\
%\hline
21&$d_{L}^{c3}$&$\stackrel{03}{[-i]}\,\stackrel{12}{(+)}|\stackrel{56}{[-]}\,\stackrel{78}{(+)}
||\stackrel{9 \;10}{[-]}\;\;\stackrel{11\;12}{(-)}\;\;\stackrel{13\;14}{[+]}$&-1&$\frac{1}{2}$&
$-\frac{1}{2}$&0&$0$&$-\frac{1}{\sqrt{3}}$&$\frac{1}{6}$&$\frac{1}{6}$&$-\frac{1}{3}$\\
%\hline
22&$d_{L}^{c3} $&$\stackrel{03}{(+i)}\,\stackrel{12}{[-]}|\stackrel{56}{[-]}\,\stackrel{78}{(+)}
||\stackrel{9 \;10}{[-]}\;\;\stackrel{11\;12}{(-)}\;\;\stackrel{13\;14}{[+]} $&-1&$-\frac{1}{2}$&
$-\frac{1}{2}$&0&$0$&$-\frac{1}{\sqrt{3}}$&$\frac{1}{6}$&$\frac{1}{6}$&$-\frac{1}{3}$\\
%\hline
23&$ u_{L}^{c3}$&$\stackrel{03}{[-i]}\,\stackrel{12}{(+)}|\stackrel{56}{(+)}\,\stackrel{78}{[-]}
||\stackrel{9 \;10}{[-]}\;\;\stackrel{11\;12}{(-)}\;\;\stackrel{13\;14}{[+]}$ &-1&$\frac{1}{2}$&
$\frac{1}{2}$&0 &$0$&$-\frac{1}{\sqrt{3}}$&$\frac{1}{6}$&$\frac{1}{6}$&$\frac{2}{3}$\\
%\hline
24&$u_{L}^{c3}$&$\stackrel{03}{(+i)}\,\stackrel{12}{[-]}|\stackrel{56}{(+)}\,\stackrel{78}{[-]}
||\stackrel{9 \;10}{[-]}\;\;\stackrel{11\;12}{(-)}\;\;\stackrel{13\;14}{[+]}$&-1&$-\frac{1}{2}$&
$\frac{1}{2}$&0&$0$&$-\frac{1}{\sqrt{3}}$&$\frac{1}{6}$&$\frac{1}{6}$&$\frac{2}{3}$\\
%\hline
\hline
25&$ \nu_{R}$&$ \stackrel{03}{(+i)}\,\stackrel{12}{(+)}|
\stackrel{56}{(+)}\,\stackrel{78}{(+)}
||\stackrel{9 \;10}{(+)}\;\;\stackrel{11\;12}{[+]}\;\;\stackrel{13\;14}{[+]} $ &1&$\frac{1}{2}$&0&
$\frac{1}{2}$&$0$&$0$&$-\frac{1}{2}$&$0$&$0$\\
%\hline 
26&$\nu_{R}$&$\stackrel{03}{[-i]}\,\stackrel{12}{[-]}|\stackrel{56}{(+)}\,\stackrel{78}{(+)}
||\stackrel{9 \;10}{(+)}\;\;\stackrel{11\;12}{[+]}\;\;\stackrel{13\;14}{[+]}$&1&$-\frac{1}{2}$&0&
$\frac{1}{2}$ &$0$&$0$&$-\frac{1}{2}$&$0$&$0$\\
%\hline
27&$e_{R}$&$\stackrel{03}{(+i)}\,\stackrel{12}{(+)}|\stackrel{56}{[-]}\,\stackrel{78}{[-]}
||\stackrel{9 \;10}{(+)}\;\;\stackrel{11\;12}{[+]}\;\;\stackrel{13\;14}{[+]}$&1&$\frac{1}{2}$&0&
$-\frac{1}{2}$&$0$&$0$&$-\frac{1}{2}$&$-1$&$-1$\\
%\hline 
28&$ e_{R} $&$\stackrel{03}{[-i]}\,\stackrel{12}{[-]}|
\stackrel{56}{[-]}\,\stackrel{78}{[-]}
||\stackrel{9 \;10}{(+)}\;\;\stackrel{11\;12}{[+]}\;\;\stackrel{13\;14}{[+]} $&1&$-\frac{1}{2}$&0&
$-\frac{1}{2}$&$0$&$0$&$-\frac{1}{2}$&$-1$&$-1$\\
%\hline
29&$e_{L}$&$\stackrel{03}{[-i]}\,\stackrel{12}{(+)}|\stackrel{56}{[-]}\,\stackrel{78}{(+)}
||\stackrel{9 \;10}{(+)}\;\;\stackrel{11\;12}{[+]}\;\;\stackrel{13\;14}{[+]}$&-1&$\frac{1}{2}$&
$-\frac{1}{2}$&0&$0$&$0$&$-\frac{1}{2}$&$-\frac{1}{2}$&$-1$\\
%\hline
30&$e_{L} $&$\stackrel{03}{(+i)}\,\stackrel{12}{[-]}|\stackrel{56}{[-]}\,\stackrel{78}{(+)}
||\stackrel{9 \;10}{(+)}\;\;\stackrel{11\;12}{[+]}\;\;\stackrel{13\;14}{[+]} $&-1&$-\frac{1}{2}$&
$-\frac{1}{2}$&0&$0$&$0$&$-\frac{1}{2}$&$-\frac{1}{2}$&$-1$\\
%\hline
31&$ \nu_{L}$&$\stackrel{03}{[-i]}\,\stackrel{12}{(+)}|\stackrel{56}{(+)}\,\stackrel{78}{[-]}
||\stackrel{9 \;10}{(+)}\;\;\stackrel{11\;12}{[+]}\;\;\stackrel{13\;14}{[+]}$ &-1&$\frac{1}{2}$&
$\frac{1}{2}$&0 &$0$&$0$&$-\frac{1}{2}$&$-\frac{1}{2}$&$0$\\
%\hline
32&$\nu_{L}$&$\stackrel{03}{(+i)}\,\stackrel{12}{[-]}|\stackrel{56}{(+)}\,\stackrel{78}{[-]}
||\stackrel{9 \;10}{(+)}\;\;\stackrel{11\;12}{[+]}\;\;\stackrel{13\;14}{[+]}$&-1&$-\frac{1}{2}$&
$\frac{1}{2}$&0&$0$&$0$&$-\frac{1}{2}$&$-\frac{1}{2}$&$0$\\
\hline%\bottomrule
%\end{tiny}
\end{tabular}} \label{Table so13+1.} 
\end{tiny}
\end{table}  
\begin{table}
\begin{tiny}
\tbl{Table~\ref{Table so13+1.} continued.}  
%\begin{tiny}
%{\begin{tabular}{|r|c||c||c|c||c|c||c|c|c||r|r|} 
 {\begin{tabular}{r c |c  |r  r | r  r | r r r | r r }
%\begin{tiny}
\hline
i&$$&$|^a\psi_i>,\,\Gamma^{(7,1)} = (-1)\,1\,, \,\Gamma^{(6)} = (1)\,-1$&$\Gamma^{(3,1)}$&$S^{12}$&
$\tau^{13}$&$\tau^{23}$&$\tau^{33}$&$\tau^{38}$&$\tau^{4}$&$Y$&$Q$\\
%\hline
%,\,\Gamma^{(7,1)} = (-1)\,1\,, \,\Gamma^{(6)} = (1)\,-1
&& ${\rm (Anti)octet}$&&&&&&&&& \\
&& ${\rm of \;(anti) quarks \;and \;(anti)leptons}$&&&&&&&&&\\
%\hline
\hline
33&$ \bar{d}_{L}^{\bar{c1}}$&$ \stackrel{03}{[-i]}\,\stackrel{12}{(+)}|
\stackrel{56}{(+)}\,\stackrel{78}{(+)}
||\stackrel{9 \;10}{[-]}\;\;\stackrel{11\;12}{[+]}\;\;\stackrel{13\;14}{[+]} $ &-1&$\frac{1}{2}$&0&
$\frac{1}{2}$&$-\frac{1}{2}$&$-\frac{1}{2\,\sqrt{3}}$&$-\frac{1}{6}$&$\frac{1}{3}$&$\frac{1}{3}$\\
%\hline 
34&$\bar{d}_{L}^{\bar{c1}}$&$\stackrel{03}{(+i)}\,\stackrel{12}{[-]}|\stackrel{56}{(+)}\,\stackrel{78}{(+)}
||\stackrel{9 \;10}{[-]}\;\;\stackrel{11\;12}{[+]}\;\;\stackrel{13\;14}{[+]}$&-1&$-\frac{1}{2}$&0&
$\frac{1}{2}$&$-\frac{1}{2}$&$-\frac{1}{2\,\sqrt{3}}$&$-\frac{1}{6}$&$\frac{1}{3}$&$\frac{1}{3}$\\
%\hline
35&$\bar{u}_{L}^{\bar{c1}}$&$\stackrel{03}{[-i]}\,\stackrel{12}{(+)}|\stackrel{56}{[-]}\,\stackrel{78}{[-]}
||\stackrel{9 \;10}{[-]}\;\;\stackrel{11\;12}{[+]}\;\;\stackrel{13\;14}{[+]}$&-1&$\frac{1}{2}$&0&
$-\frac{1}{2}$&$-\frac{1}{2}$&$-\frac{1}{2\,\sqrt{3}}$&$-\frac{1}{6}$&$-\frac{2}{3}$&$-\frac{2}{3}$\\
%\hline
36&$ \bar{u}_{L}^{\bar{c1}} $&$\stackrel{03}{(+i)}\,\stackrel{12}{[-]}|
\stackrel{56}{[-]}\,\stackrel{78}{[-]}
||\stackrel{9 \;10}{[-]}\;\;\stackrel{11\;12}{[+]}\;\;\stackrel{13\;14}{[+]} $&-1&$-\frac{1}{2}$&0&
$-\frac{1}{2}$&$-\frac{1}{2}$&$-\frac{1}{2\,\sqrt{3}}$&$-\frac{1}{6}$&$-\frac{2}{3}$&$-\frac{2}{3}$\\
%\hline
37&$\bar{d}_{R}^{\bar{c1}}$&$\stackrel{03}{(+i)}\,\stackrel{12}{(+)}|\stackrel{56}{(+)}\,\stackrel{78}{[-]}
||\stackrel{9 \;10}{[-]}\;\;\stackrel{11\;12}{[+]}\;\;\stackrel{13\;14}{[+]}$&1&$\frac{1}{2}$&
$\frac{1}{2}$&0&$-\frac{1}{2}$&$-\frac{1}{2\,\sqrt{3}}$&$-\frac{1}{6}$&$-\frac{1}{6}$&$\frac{1}{3}$\\
%\hline
38&$\bar{d}_{R}^{\bar{c1}} $&$\stackrel{03}{[-i]}\,\stackrel{12}{[-]}|\stackrel{56}{(+)}\,\stackrel{78}{[-]}
||\stackrel{9 \;10}{[-]}\;\;\stackrel{11\;12}{[+]}\;\;\stackrel{13\;14}{[+]} $&1&$-\frac{1}{2}$&
$\frac{1}{2}$&0&$-\frac{1}{2}$&$-\frac{1}{2\,\sqrt{3}}$&$-\frac{1}{6}$&$-\frac{1}{6}$&$\frac{1}{3}$\\
%\hline
39&$ \bar{u}_{R}^{\bar{c1}}$&$\stackrel{03}{(+i)}\,\stackrel{12}{(+)}|\stackrel{56}{[-]}\,\stackrel{78}{(+)}
||\stackrel{9 \;10}{[-]}\;\;\stackrel{11\;12}{[+]}\;\;\stackrel{13\;14}{[+]}$ &1&$\frac{1}{2}$&
$-\frac{1}{2}$&0 &$-\frac{1}{2}$&$-\frac{1}{2\,\sqrt{3}}$&$-\frac{1}{6}$&$-\frac{1}{6}$&$-\frac{2}{3}$\\
%\hline
40&$\bar{u}_{R}^{\bar{c1}}$&$\stackrel{03}{[-i]}\,\stackrel{12}{[-]}|\stackrel{56}{[-]}\,\stackrel{78}{(+)}
||\stackrel{9 \;10}{[-]}\;\;\stackrel{11\;12}{[+]}\;\;\stackrel{13\;14}{[+]}$
%\stackrel{9 \;10}{[-]}\;\;\stackrel{11\;12}{[+]}\;\;\stackrel{13\;14}{[+]}$
&1&$-\frac{1}{2}$&
$-\frac{1}{2}$&0&$-\frac{1}{2}$&$-\frac{1}{2\,\sqrt{3}}$&$-\frac{1}{6}$&$-\frac{1}{6}$&$-\frac{2}{3}$\\
%\hline
\hline
41&$ \bar{d}_{L}^{\bar{c2}}$&$ \stackrel{03}{[-i]}\,\stackrel{12}{(+)}|
\stackrel{56}{(+)}\,\stackrel{78}{(+)}
||\stackrel{9 \;10}{(+)}\;\;\stackrel{11\;12}{(-)}\;\;\stackrel{13\;14}{[+]} $ 
&-1&$\frac{1}{2}$&0&
$\frac{1}{2}$&$\frac{1}{2}$&$-\frac{1}{2\,\sqrt{3}}$&$-\frac{1}{6}$&$\frac{1}{3}$&$\frac{1}{3}$\\
%\hline
42&$\bar{d}_{L}^{\bar{c2}}$&$\stackrel{03}{(+i)}\,\stackrel{12}{[-]}|\stackrel{56}{(+)}\,\stackrel{78}{(+)}
||\stackrel{9 \;10}{(+)}\;\;\stackrel{11\;12}{(-)}\;\;\stackrel{13\;14}{[+]}$
&-1&$-\frac{1}{2}$&0&
$\frac{1}{2}$&$\frac{1}{2}$&$-\frac{1}{2\,\sqrt{3}}$&$-\frac{1}{6}$&$\frac{1}{3}$&$\frac{1}{3}$\\
%\hline
43&$\bar{u}_{L}^{\bar{c2}}$&$\stackrel{03}{[-i]}\,\stackrel{12}{(+)}|\stackrel{56}{[-]}\,\stackrel{78}{[-]}
||\stackrel{9 \;10}{(+)}\;\;\stackrel{11\;12}{(-)}\;\;\stackrel{13\;14}{[+]}$
&-1&$\frac{1}{2}$&0&
$-\frac{1}{2}$&$\frac{1}{2}$&$-\frac{1}{2\,\sqrt{3}}$&$-\frac{1}{6}$&$-\frac{2}{3}$&$-\frac{2}{3}$\\
%\hline
44&$ \bar{u}_{L}^{\bar{c2}} $&$\stackrel{03}{(+i)}\,\stackrel{12}{[-]}|
\stackrel{56}{[-]}\,\stackrel{78}{[-]}
||\stackrel{9 \;10}{(+)}\;\;\stackrel{11\;12}{(-)}\;\;\stackrel{13\;14}{[+]} $
&-1&$-\frac{1}{2}$&0&
$-\frac{1}{2}$&$\frac{1}{2}$&$-\frac{1}{2\,\sqrt{3}}$&$-\frac{1}{6}$&$-\frac{2}{3}$&$-\frac{2}{3}$\\
%\hline
45&$\bar{d}_{R}^{\bar{c2}}$&$\stackrel{03}{(+i)}\,\stackrel{12}{(+)}|\stackrel{56}{(+)}\,\stackrel{78}{[-]}
||\stackrel{9 \;10}{(+)}\;\;\stackrel{11\;12}{(-)}\;\;\stackrel{13\;14}{[+]}$
&1&$\frac{1}{2}$&
$\frac{1}{2}$&0&$\frac{1}{2}$&$-\frac{1}{2\,\sqrt{3}}$&$-\frac{1}{6}$&$-\frac{1}{6}$&$\frac{1}{3}$\\
%\hline
46&$\bar{d}_{R}^{\bar{c2}} $&$\stackrel{03}{[-i]}\,\stackrel{12}{[-]}|\stackrel{56}{(+)}\,\stackrel{78}{[-]}
||\stackrel{9 \;10}{(+)}\;\;\stackrel{11\;12}{(-)}\;\;\stackrel{13\;14}{[+]} $
&1&$-\frac{1}{2}$&
$\frac{1}{2}$&0&$\frac{1}{2}$&$-\frac{1}{2\,\sqrt{3}}$&$-\frac{1}{6}$&$-\frac{1}{6}$&$\frac{1}{3}$\\
%\hline
47&$ \bar{u}_{R}^{\bar{c2}}$&$\stackrel{03}{(+i)}\,\stackrel{12}{(+)}|\stackrel{56}{[-]}\,\stackrel{78}{(+)}
||\stackrel{9 \;10}{(+)}\;\;\stackrel{11\;12}{(-)}\;\;\stackrel{13\;14}{[+]}$
 &1&$\frac{1}{2}$&
$-\frac{1}{2}$&0 &$\frac{1}{2}$&$-\frac{1}{2\,\sqrt{3}}$&$-\frac{1}{6}$&$-\frac{1}{6}$&$-\frac{2}{3}$\\
%\hline
48&$\bar{u}_{R}^{\bar{c2}}$&$\stackrel{03}{[-i]}\,\stackrel{12}{[-]}|\stackrel{56}{[-]}\,\stackrel{78}{(+)}
||\stackrel{9 \;10}{(+)}\;\;\stackrel{11\;12}{(-)}\;\;\stackrel{13\;14}{[+]}$
&1&$-\frac{1}{2}$&
$-\frac{1}{2}$&0&$\frac{1}{2}$&$-\frac{1}{2\,\sqrt{3}}$&$-\frac{1}{6}$&$-\frac{1}{6}$&$-\frac{2}{3}$\\
%\hline
\hline
%\hline 
%$\cdots$ &&&&&&&&&&& \\
%\hline\hline
49&$ \bar{d}_{L}^{\bar{c3}}$&$ \stackrel{03}{[-i]}\,\stackrel{12}{(+)}|
\stackrel{56}{(+)}\,\stackrel{78}{(+)}
||\stackrel{9 \;10}{(+)}\;\;\stackrel{11\;12}{[+]}\;\;\stackrel{13\;14}{(-)} $ &-1&$\frac{1}{2}$&0&
$\frac{1}{2}$&$0$&$\frac{1}{\sqrt{3}}$&$-\frac{1}{6}$&$\frac{1}{3}$&$\frac{1}{3}$\\
%\hline 
50&$\bar{d}_{L}^{\bar{c3}}$&$\stackrel{03}{(+i)}\,\stackrel{12}{[-]}|\stackrel{56}{(+)}\,\stackrel{78}{(+)}
||\stackrel{9 \;10}{(+)}\;\;\stackrel{11\;12}{[+]}\;\;\stackrel{13\;14}{(-)} $&-1&$-\frac{1}{2}$&0&
$\frac{1}{2}$&$0$&$\frac{1}{\sqrt{3}}$&$-\frac{1}{6}$&$\frac{1}{3}$&$\frac{1}{3}$\\
%\hline
51&$\bar{u}_{L}^{\bar{c3}}$&$\stackrel{03}{[-i]}\,\stackrel{12}{(+)}|\stackrel{56}{[-]}\,\stackrel{78}{[-]}
||\stackrel{9 \;10}{(+)}\;\;\stackrel{11\;12}{[+]}\;\;\stackrel{13\;14}{(-)} $&-1&$\frac{1}{2}$&0&
$-\frac{1}{2}$&$0$&$\frac{1}{\sqrt{3}}$&$-\frac{1}{6}$&$-\frac{2}{3}$&$-\frac{2}{3}$\\
%\hline
52&$ \bar{u}_{L}^{\bar{c3}} $&$\stackrel{03}{(+i)}\,\stackrel{12}{[-]}|
\stackrel{56}{[-]}\,\stackrel{78}{[-]}
||\stackrel{9 \;10}{(+)}\;\;\stackrel{11\;12}{[+]}\;\;\stackrel{13\;14}{(-)}  $&-1&$-\frac{1}{2}$&0&
$-\frac{1}{2}$&$0$&$\frac{1}{\sqrt{3}}$&$-\frac{1}{6}$&$-\frac{2}{3}$&$-\frac{2}{3}$\\
%\hline
53&$\bar{d}_{R}^{\bar{c3}}$&$\stackrel{03}{(+i)}\,\stackrel{12}{(+)}|\stackrel{56}{(+)}\,\stackrel{78}{[-]}
||\stackrel{9 \;10}{(+)}\;\;\stackrel{11\;12}{[+]}\;\;\stackrel{13\;14}{(-)} $&1&$\frac{1}{2}$&
$\frac{1}{2}$&0&$0$&$\frac{1}{\sqrt{3}}$&$-\frac{1}{6}$&$-\frac{1}{6}$&$\frac{1}{3}$\\
%\hline
54&$\bar{d}_{R}^{\bar{c3}} $&$\stackrel{03}{[-i]}\,\stackrel{12}{[-]}|\stackrel{56}{(+)}\,\stackrel{78}{[-]}
||\stackrel{9 \;10}{(+)}\;\;\stackrel{11\;12}{[+]}\;\;\stackrel{13\;14}{(-)} $&1&$-\frac{1}{2}$&
$\frac{1}{2}$&0&$0$&$\frac{1}{\sqrt{3}}$&$-\frac{1}{6}$&$-\frac{1}{6}$&$\frac{1}{3}$\\
%\hline
55&$ \bar{u}_{R}^{\bar{c3}}$&$\stackrel{03}{(+i)}\,\stackrel{12}{(+)}|\stackrel{56}{[-]}\,\stackrel{78}{(+)}
||\stackrel{9 \;10}{(+)}\;\;\stackrel{11\;12}{[+]}\;\;\stackrel{13\;14}{(-)} $ &1&$\frac{1}{2}$&
$-\frac{1}{2}$&0 &$0$&$\frac{1}{\sqrt{3}}$&$-\frac{1}{6}$&$-\frac{1}{6}$&$-\frac{2}{3}$\\
%\hline
56&$\bar{u}_{R}^{\bar{c3}}$&$\stackrel{03}{[-i]}\,\stackrel{12}{[-]}|\stackrel{56}{[-]}\,\stackrel{78}{(+)}
||\stackrel{9 \;10}{(+)}\;\;\stackrel{11\;12}{[+]}\;\;\stackrel{13\;14}{(-)} $&1&$-\frac{1}{2}$&
$-\frac{1}{2}$&0&$0$&$\frac{1}{\sqrt{3}}$&$-\frac{1}{6}$&$-\frac{1}{6}$&$-\frac{2}{3}$\\
%\hline
\hline
%$\cdots$ &&&&&&&&&&& \\
%\hline\hline
57&$ \bar{e}_{L}$&$ \stackrel{03}{[-i]}\,\stackrel{12}{(+)}|
\stackrel{56}{(+)}\,\stackrel{78}{(+)}
||\stackrel{9 \;10}{[-]}\;\;\stackrel{11\;12}{(-)}\;\;\stackrel{13\;14}{(-)} $ &-1&$\frac{1}{2}$&0&
$\frac{1}{2}$&$0$&$0$&$\frac{1}{2}$&$1$&$1$\\
%\hline 
58&$\bar{e}_{L}$&$\stackrel{03}{(+i)}\,\stackrel{12}{[-]}|\stackrel{56}{(+)}\,\stackrel{78}{(+)}
||\stackrel{9 \;10}{[-]}\;\;\stackrel{11\;12}{(-)}\;\;\stackrel{13\;14}{(-)}$&-1&$-\frac{1}{2}$&0&
$\frac{1}{2}$ &$0$&$0$&$\frac{1}{2}$&$1$&$1$\\
%\hline
59&$\bar{\nu}_{L}$&$\stackrel{03}{[-i]}\,\stackrel{12}{(+)}|\stackrel{56}{[-]}\,\stackrel{78}{[-]}
||\stackrel{9 \;10}{[-]}\;\;\stackrel{11\;12}{(-)}\;\;\stackrel{13\;14}{(-)}$&-1&$\frac{1}{2}$&0&
$-\frac{1}{2}$&$0$&$0$&$\frac{1}{2}$&$0$&$0$\\
%\hline 
60&$ \bar{\nu}_{L} $&$\stackrel{03}{(+i)}\,\stackrel{12}{[-]}|
\stackrel{56}{[-]}\,\stackrel{78}{[-]}
||\stackrel{9 \;10}{[-]}\;\;\stackrel{11\;12}{(-)}\;\;\stackrel{13\;14}{(-)} $&-1&$-\frac{1}{2}$&0&
$-\frac{1}{2}$&$0$&$0$&$\frac{1}{2}$&$0$&$0$\\
%\hline
61&$\bar{\nu}_{R}$&$\stackrel{03}{(+i)}\,\stackrel{12}{(+)}|\stackrel{56}{[-]}\,\stackrel{78}{(+)}
||\stackrel{9 \;10}{[-]}\;\;\stackrel{11\;12}{(-)}\;\;\stackrel{13\;14}{(-)}$&1&$\frac{1}{2}$&
$-\frac{1}{2}$&0&$0$&$0$&$\frac{1}{2}$&$\frac{1}{2}$&$0$\\
%\hline
62&$\bar{\nu}_{R} $&$\stackrel{03}{[-i]}\,\stackrel{12}{[-]}|\stackrel{56}{[-]}\,\stackrel{78}{(+)}
||\stackrel{9 \;10}{[-]}\;\;\stackrel{11\;12}{(-)}\;\;\stackrel{13\;14}{(-)} $&1&$-\frac{1}{2}$&
$-\frac{1}{2}$&0&$0$&$0$&$\frac{1}{2}$&$\frac{1}{2}$&$0$\\
%\hline
63&$ \bar{e}_{R}$&$\stackrel{03}{(+i)}\,\stackrel{12}{(+)}|\stackrel{56}{(+)}\,\stackrel{78}{[-]}
||\stackrel{9 \;10}{[-]}\;\;\stackrel{11\;12}{(-)}\;\;\stackrel{13\;14}{(-)}$ &1&$\frac{1}{2}$&
$\frac{1}{2}$&0 &$0$&$0$&$\frac{1}{2}$&$\frac{1}{2}$&$1$\\
%\hline
64&$\bar{e}_{R}$&$\stackrel{03}{[-i]}\,\stackrel{12}{[-]}|\stackrel{56}{(+)}\,\stackrel{78}{[-]}
||\stackrel{9 \;10}{[-]}\;\;\stackrel{11\;12}{(-)}\;\;\stackrel{13\;14}{(-)}$&1&$-\frac{1}{2}$&
$\frac{1}{2}$&0&$0$&$0$&$\frac{1}{2}$&$\frac{1}{2}$&$1$\\
\hline
%\end{tiny}
\end{tabular}} \label{Table so13+1.pt2} 
\end{tiny}
\end{table} 

In Table~\ref{Table so13+1.} the starting state is chosen
to be $ \stackrel{03}{(+i)}\,\stackrel{12}{(+)}|\stackrel{56}{(+)}\,\stackrel{78}{(+)}
||\stackrel{9 \;10}{(+)}\;\;\stackrel{11\;12}{(-)}\;\;\stackrel{13\;14}{(-)} $. We could make 
any other choice of products of nilpotents and projectors, let say the state 
$\stackrel{03}{[-i]}\,\stackrel{12}{(+)}|\stackrel{56}{(+)}\,\stackrel{78}{[-]}
||\stackrel{9 \;10}{(+)}\;\;\stackrel{11\;12}{(-)}\;\;\stackrel{13\;14}{(-)}$, which 
is the state in the seventh line of Table~\ref{Table so13+1.}. All the states of 
one representation can be obtained from the starting state by applying on the starting state the 
generators  $S^{ab}$.  From the first state, for example,  we
obtain the seventh one by the application of $S^{0\,7}$ (or of $S^{0\,8}$,  $S^{3\,7}$,  
$S^{3\,8}$).
% 10.04.2017 at 22:00

Let us make a few examples to get inside how can one read the quantum numbers of states from 
$7$ products of nilpotents and projectors. All nilpotents and projectors are eigen states, 
Eq.~(\ref{grapheigen}), of  Cartan sub-algebra, Eq.~(\ref{choicecartan}). 

\vspace{2mm}

{\it Example 1.}: Let us calculate properties of  the two states: The first state ---
$ \stackrel{03}{(+i)}\stackrel{12}{(+)}|\stackrel{56}{(+)}\stackrel{78}{(+)}||%
\stackrel{9 \;10}{(+)}\stackrel{11\;12}{(-)}\stackrel{13\;14}{(-)}|\psi_{0} \rangle$ --- and 
the seventh  state --- $\stackrel{03}{[-i]}\,\stackrel{12}{(+)}|\stackrel{56}{(+)}%
\stackrel{78}{[-]}||\stackrel{9 \;10}{(+)}\stackrel{11\;12}{(-)}\stackrel{13\;14}{(-)}|\psi_{0}%
\rangle$ ---  of Table~\ref{Table so13+1.}. 

The handedness of the whole  one Weyl representation ($64$ states) follows from Eqs.~(\ref{hand},%
\ref{choicecartan}): $\Gamma^{(14)}=$  $ i^{13} 2^7  S^{03}  S^{12} \cdots S^{13\,14}$. 
This operator  gives, when applied on the first state of Table~\ref{Table so13+1.}, the eigenvalue 
$=i^{13} 2^{7} \frac{i}{2}  (\frac{1}{2})^4   (-\frac{1}{2})^2 =-1$ (since the operator 
$S^{03}$ applied on the nilpotent $ \stackrel{03}{(+i)}$ gives  the eigenvalue  $\frac{k}{2}= 
\frac{i}{2} $, the rest four   operators 
have the eigenvalues  $\frac{1}{2}$, and the last two  $ - \frac{1}{2}$, Eq.~(\ref{grapheigen})). 

In an equivalent way we calculate the  handedness $\Gamma^{(3,1)}$ of  these two states   
in $d=(3+1)$:  The operator $\Gamma^{(3,1)} = i^3 2^2 S^{03} S^{12}$, applied on the first state, 
gives $1$ --- the right handedness, while $\Gamma^{(3,1)}$ is for the seventh state $-1$ ---
 the left handedness. 

The weak charge operator $\tau^{13} (=\frac{1}{2}\, (S^{56}-S^{78}))$, Eq.~(\ref{so42}), 
applied on the first state, gives  the eigenvalue $0$: $\frac{1}{2}\, (\frac{1}{2}-\frac{1}{2})$,  
The eigenvalue of $\tau^{13}$  is for the seventh state $\frac{1}{2}$: $\frac{1}{2} 
(\frac{1}{2}- (-\frac{1}{2}))$,  $\tau^{23}$ $(=\frac{1}{2}(S^{56} + S^{78}))$, applied on  
the first state, gives as its eigenvalue $\frac{1}{2}$, while when  $\tau^{23}$ applies on the seventh 
state gives $0$. The "fermion charge" operator $\tau^4$ ($= - \frac{1}{3} (S^{9\,10}+S^{11\,12}
+S^{13\,14})$, Eq.~(\ref{so64})) gives, when applied on any of these two states, the eigenvalues 
$- \frac{1}{3} ( \frac{1}{2} - \frac{1}{2} - \frac{1}{2}) = \frac{1}{6}$. 
Correspondingly is the hyper charge $Y$ $(=\tau^{23} + \tau^{4})$ of these two 
states $Y= (\frac{2}{3}, \frac{1}{6}) $, respectively, what the {\it standard model} assumes for 
$u_{R}$ and $u_{L}$, respectively. 

One finds for the colour charge of these two states, 
%(all the operators $S^{ab}$ appearing in the expressions for the colour charge are diagonal),  
($\tau^{3 3}$, $ \tau^{3 8}$) $(=(\frac{1}{2} (S^{9\,10} - S^{11\,12})$,  
$\frac{1}{\sqrt{3}} (S^{9\,10} + S^{11\,12} - 2 S^{13\,14}))$ the
eigenvalues $ (1/2,1/(2\sqrt{3})$).

The first and the seventh states differ in the handedness $\Gamma^{(3,1)}=$ ($1,-1$),
in the weak charge $\tau^{13}=$($0, \frac{1}{2}$) and the hyper charge 
$Y=$  ($\frac{2}{3},\frac{1}{6}$), respectively.
All the states of this octet --- $SO(7,1)$ --- have the same colour charge and the same 
"fermion charge" (the difference in the hyper charge $Y$ is caused by the difference in
 $\tau^{23}=$ ($\frac{1}{2},0$)).

The states for the $d_{R}$-quark and $d_{L}$-quark of the same octet follow from the state 
 $u_{R}$ and $u_{L}$, respectively, by the application of $S^{5 7}$ (or $S^{5 8}$, $S^{6 7}$,
 $S^{6 8}$).

All  the $SO(7,1)$ ($\Gamma^{(7,1)}= 1$) part of the $SO(13,1)$ spinor representation are the
 same for either quarks of all the three colours (quarks states appear in 
Table~\ref{Table so13+1.} from the first to the $24^{th}$ line) or for the colourless leptons  
(leptons appear in Table~\ref{Table so13+1.} from the $25^{th}$ line to the $32^{nd}$ line). 

Leptons distinguish  from quarks in the part represented by nilpotents and projectors, which is 
determined by the eigenstates of the Cartan subalgebra of ($S^{9\,10}, S^{11\,12}, S^{13\,14}$). 
Taking into account Eq.~(\ref{grapheigen}) one calculates that ($\tau^{3 3}$, $\tau^{3 8}$) is for 
the colourless part of the lepton states $(\nu_{R,L}, e_{R,L})$ ---  
($\cdots||\stackrel{9 \;10}{(+)}\;\;\stackrel{11\;12}{[+]}\;\;\stackrel{13\;14}{[+]} $) --- 
equal to $= (0, 0)$, while the "fermion charge" $\tau^{4}$ is for these states equal to
 $-\frac{1}{2}$ (just as assumed by the {\it standard model}).

 Let us point out that the octet $SO(7,1)$ manifests how the spin and the weak and hyper charges 
are related.

\vspace{2mm}

{\it Example 2.}: Let us look at the properties of  the anti-quark and anti-lepton states of one 
fundamental representation of the $SO(13,1)$ group. These states are presented in 
Table~\ref{Table so13+1.} from the $33^{rd}$ line to the $64^{th}$ line, representing anti-quarks 
(the first three octets) and anti-leptons (the last octet). 

Again, all the anti-octets, the $ SO(7,1)$ ($\Gamma^{(7,1)}= - 1$) part of the $SO(13,1)$ 
representation, are the same either for anti-quarks or for anti-leptons. The last three products 
of nilpotents and projectors (the part appearing in Table~\ref{Table so13+1.} after "$||$") 
determine anti-colours for the anti-quarks states and the anti-colourless state for anti-leptons.  

Let us add that all the anti-spinor states are reachable from the spinor states (and opposite) 
 by the application of the operator~\cite{discretesym} 
$\mathbb{C}_{{\cal N}}  {\cal P}_{{\cal N}}$~\footnote{Discrete symmetries in $d=(3+1)$ follow
 from the corresponding definition of these symmetries in $d$-
dimensional space~\cite{discretesym}.  
This operator is defined as: $\mathbb{C}_{{\cal N}}  {\cal P}_{{\cal N}} $ $=\gamma^0
 \prod_{\Im \gamma^s, s=5}^{d} \gamma^s\,\,I_{\vec{x}_3} \,I_{x^6,x^8,\dots,x^{d}}\,$,
 where $\gamma^0$ and $\gamma^1$ are real, 
$\gamma^2$ imaginary, $\gamma^3$ real, $\gamma^5$ imaginary, $\gamma^6$ real, alternating 
imaginary and real up to $\gamma^d$, which is in even dimensional spaces real.  
$\gamma^a$'s appear in the ascending order.
Operators $I$ operate as follows: $\quad I_{x^0} x^0 = -x^0\,$; $
I_{x} x^a =- x^a\,$; $  I_{x^0} x^a = (-x^0,\vec{x})\,$; $ I_{\vec{x}} \vec{x} = -\vec{x}
\,$; $I_{\vec{x}_{3}} x^a = (x^0, -x^1,-x^2,-x^3,x^5, x^6,\dots, x^d)\,$; 
$I_{x^5,x^7,\dots,x^{d-1}}$ $(x^0,x^1,x^2,x^3,x^5,x^6,x^7,x^8,
\dots,x^{d-1},x^d)$ $=(x^0,x^1,x^2,x^3,-x^5,x^6,-x^7,\dots,-x^{d-1},x^d)$;
 $I_{x^6,x^8,\dots,x^d}$ 
$(x^0,x^1,x^2,x^3,x^5,x^6,x^7,x^8,\dots,$ $x^{d-1},x^d)$
$=(x^0,x^1,x^2,x^3,x^5,-x^6,x^7,-x^8,\dots$ $,x^{d-1},-x^d)$, $d=2n$.}.
The part of this operator,
 which operates on only the spinor part of the state (presented in Table~\ref{Table so13+1.}), 
is $\mathbb{C}_{{\cal N}}  {\cal P}_{{\cal N}}|_{spinor}$ 
$=\gamma^0  \prod_{\Im \gamma^s, s=5}^{d} \gamma^s$. 
Taking into account 
Eq.~(\ref{snmb:gammatildegamma}) and this operator  one finds that
 $\mathbb{C}_{{\cal N}}  {\cal P}_{{\cal N}}|_{spinor}$
transforms $u^{c1}_{R}$ from the first line of Table~\ref{Table so13+1.} into 
${\bar u}^{\bar{c1}}_{L}$ from the $35^{th}$ line of the same table. When the operator 
$\mathbb{C}_{{\cal N}}  {\cal P}_{{\cal N}}|_{spinor}$ applies on $\nu_{R}$ (the 
$25^{th}$ line of the same table,  with the colour chargeless part  equal to 
$\cdots||\stackrel{9 \;10}{(+)}\;\;\stackrel{11\;12}{[+]}\;\;\stackrel{13\;14}{[+]} $),  
transforms  $\nu_{R}$ into 
${\bar \nu}_{L}$  (the $59^{th}$ line of the table, with the colour anti-chargeless part equal to
($\cdots||\stackrel{9 \;10}{[-]}\;\;\stackrel{11\;12}{(-)}\;\;\stackrel{13\;14}{(-)} $).

\subsection{A short inside into families in the {\it spin-charge-family} theory}
\label{families}

The operators $\tilde{S}^{ab}$, commuting with $S^{ab}$ (Eq.~(\ref{sabtildesab})),
 transform any spinor state, presented in Table~\ref{Table so13+1.}, to the same state of 
another family,  orthogonal to the starting state and correspondingly to all the states of the 
starting family.

Applying the opeartor $\tilde{S}^{01}$ ($= \frac{i}{2} \gamma^0 \gamma^1$), 
for example, on $\nu_{R}$ (the $25^{th}$ line of Table~\ref{Table so13+1.} and the last line on 
Table~\ref{Table III.}), one obtains, taking into account 
Eq.~(\ref{snmb:gammatildegamma}),  the $\nu_{R7}$ state belonging to another family, presented 
in the seventh  line of Table~\ref{Table III.}. 

Operators $S^{ab}$ transform  $\nu_{R}$ (the $25^{th}$ line of Table~\ref{Table so13+1.}, 
presented in Table~\ref{Table III.} in the eighth line, carrying the name $\nu_{R 8}$) into all the 
rest of the $64$ states of this eighth family,  presented in Table~\ref{Table so13+1.}. The operator
$S^{11\,13}$, for example, transforms $\nu_{R 8}$ into $u_{R 8}$ (presented in the first line 
of Table~\ref{Table so13+1.}), while it transforms $\nu_{R 7}$ into $u_{R 7}$.

Table~\ref{Table III.} represents eight families of neutrinos, which distinguish among themselves 
in the family quantum numbers: ($\tilde{\tau}^{13}$, $\tilde{N}_{L}$, $\tilde{\tau}^{23}$, 
$\tilde{N}_{R}$, $\tilde{\tau}^{4}$). These family quantum numbers can be expressed by 
$\tilde{S}^{ab}$ as presented in Eqs.~(\ref{so1+3}, \ref{so42}, \ref{so64}), if $S^{ab}$ are 
replaced by $\tilde{S}^{ab}$.

Eight families decouples into two groups of four families, one ($I$) is a doublet with respect to 
($\vec{\tilde{N}}_{L}$ and  $\vec{\tilde{\tau}}^{1}$) and  a singlet with respect to 
($\vec{\tilde{N}}_{R}$ and  $\vec{\tilde{\tau}}^{2}$), the other ($II$) is a singlet with respect to 
($\vec{\tilde{N}}_{L}$ and  $\vec{\tilde{\tau}}^{1}$) and  a doublet with with respect to 
($\vec{\tilde{N}}_{R}$ and  $\vec{\tilde{\tau}}^{2}$).

All the families follow from the starting one by the application of the operators 
($\tilde{N}^{\pm}_{R,L}$, $\tilde{\tau}^{(2,1)\pm}$), Eq.~(\ref{plusminus}).  The generators 
($N^{\pm}_{R,L} $, $\tau^{(2,1)\pm}$), Eq.~(\ref{plusminus}), transform $\nu_{R1}$
 to all the members belonging to the $SO(7,1)$ group of one family, $S^{st}, (s,t) =(9\cdots,14)$
transform quarks of one colour to quarks of other colours or to leptons. 

\begin{table}
\begin{tiny}
% \begin{center}
\tbl{Eight families of the right handed  neutrino $\nu_{R}$ (appearing in the $25^{th}$ line of 
Table~\ref{Table so13+1.}), with spin $\frac{1}{2}$. % colour chargeless  $(\tau^{33}=0$, $\tau^{38}=0)$. 
$\nu_{R i}, i=(1,\cdots,8)$, carries the family quantum numbers $\tilde{\tau}^{13}$, 
$\tilde{N}^{3}_{L}$, $\tilde{\tau}^{23}$, $\tilde{N}^{3}_{R}$ and $\tilde{\tau}^{4}$. 
Eight families decouple into two groups of four families.}
{\begin{tabular}{|c|c|c|r r r r r|}
 \hline
 &&&$\tilde{\tau}^{13}$&$\tilde{\tau}^{23}$&$\tilde{N}_{L}^{3}$&$\tilde{N}_{R}^{3}$&$\tilde{\tau}^{4}$\\
 \hline
 $I$&
%$u^{c1}_{R\,1}$& $ \stackrel{03}{(+i)}\,\stackrel{12}{[+]}|\stackrel{56}{[+]}\,
%\stackrel{78}{(+)} ||   \stackrel{9 \;10}{(+)}\;\;\stackrel{11\;12}{(-)}\;\;
%\stackrel{13\;14}{(-)}$ & 
   $\nu_{R\,1}$&
   $ \stackrel{03}{(+i)}\,\stackrel{12}{[+]}|\stackrel{56}{[+]}\,\stackrel{78}{(+)} ||
   \stackrel{9 \;10}{(+)}\;\;\stackrel{11\;12}{[+]}\;\;\stackrel{13\;14}{[+]}$ 
  &$-\frac{1}{2}$&$0$&$-\frac{1}{2}$&$0$&$-\frac{1}{2}$ 
 \\
  $I$&
%$u^{c1}_{R\,2}$&   $ \stackrel{03}{[+i]}\,\stackrel{12}{(+)}|\stackrel{56}{[+]}\,
%\stackrel{78}{(+)} ||   \stackrel{9 \;10}{(+)}\;\;\stackrel{11\;12}{(-)}\;\;
%\stackrel{13\;14}{(-)}$ & 
   $\nu_{R\,2}$&
   $ \stackrel{03}{[+i]}\,\stackrel{12}{(+)}|\stackrel{56}{[+]}\,\stackrel{78}{(+)} ||
   \stackrel{9 \;10}{(+)}\;\;\stackrel{11\;12}{[+]}\;\;\stackrel{13\;14}{[+]}$ 
  &$-\frac{1}{2}$&$0$&$\frac{1}{2}$&$0$&$-\frac{1}{2}$
 \\
  $I$&
%$u^{c1}_{R\,3}$&   $ \stackrel{03}{(+i)}\,\stackrel{12}{[+]}|\stackrel{56}{(+)}\,
%\stackrel{78}{[+]} ||   \stackrel{9 \;10}{(+)}\;\;\stackrel{11\;12}{(-)}\;\;
%\stackrel{13\;14}{(-)}$ & 
   $\nu_{R\,3}$&
   $ \stackrel{03}{(+i)}\,\stackrel{12}{[+]}|\stackrel{56}{(+)}\,\stackrel{78}{[+]} ||
   \stackrel{9 \;10}{(+)}\;\;\stackrel{11\;12}{[+]}\;\;\stackrel{13\;14}{[+]}$ 
  &$\frac{1}{2}$&$0$&$-\frac{1}{2}$&$0$&$-\frac{1}{2}$
 \\
 $I$&
%$u^{c1}_{R\,4}$&  $ \stackrel{03}{[+i]}\,\stackrel{12}{(+)}|\stackrel{56}{(+)}\,
%\stackrel{78}{[+]} ||   \stackrel{9 \;10}{(+)}\;\;\stackrel{11\;12}{(-)}\;\;
%\stackrel{13\;14}{(-)}$ & 
  $\nu_{R\,4}$&
  $ \stackrel{03}{[+i]}\,\stackrel{12}{(+)}|\stackrel{56}{(+)}\,\stackrel{78}{[+]} ||
  \stackrel{9 \;10}{(+)}\;\;\stackrel{11\;12}{[+]}\;\;\stackrel{13\;14}{[+]}$ 
  &$\frac{1}{2}$&$0$&$\frac{1}{2}$&$0$&$-\frac{1}{2}$
  \\
  \hline
  $II$& 
%$u^{c1}_{R\,5}$&        $ \stackrel{03}{[+i]}\,\stackrel{12}{[+]}|\stackrel{56}{[+]}\,
%\stackrel{78}{[+]}|| \stackrel{9 \;10}{(+)}\;\;\stackrel{11\;12}{(-)}\;\;
%\stackrel{13\;14}{(-)}$ & 
  $\nu_{R\,5}$&  $ \stackrel{03}{[+i]}\,\stackrel{12}{[+]}|\stackrel{56}{[+]}\,
\stackrel{78}{[+]}|| \stackrel{9 \;10}{(+)}\;\;\stackrel{11\;12}{[+]}\;\;
\stackrel{13\;14}{[+]}$ 
  &$0$&$-\frac{1}{2}$&$0$&$-\frac{1}{2}$&$-\frac{1}{2}$
 \\ 
  $II$& 
%$u^{c1}_{R\,6}$& $ \stackrel{03}{(+i)}\,\stackrel{12}{(+)}|\stackrel{56}{[+]}\,
%\stackrel{78}{[+]}|| \stackrel{9 \;10}{(+)}\;\;\stackrel{11\;12}{(-)}\;\;
%\stackrel{13\;14}{(-)}$ &  
$\nu_{R\,6}$& $ \stackrel{03}{(+i)}\,\stackrel{12}{(+)}|\stackrel{56}{[+]}\,
\stackrel{78}{[+]}||       \stackrel{9 \;10}{(+)}\;\;\stackrel{11\;12}{[+]}\;\;
\stackrel{13\;14}{[+]}$       &$0$&$-\frac{1}{2}$&$0$&$\frac{1}{2}$&$-\frac{1}{2}$
 \\ 
 $II$& 
%$u^{c1}_{R\,7}$& $ \stackrel{03}{[+i]}\,\stackrel{12}{[+]}|\stackrel{56}{(+)}\,
%\stackrel{78}{(+)}|| \stackrel{9 \;10}{(+)}\;\;\stackrel{11\;12}{(-)}\;\;
%\stackrel{13\;14}{(-)}$ &  
$\nu_{R\,7}$&
$ \stackrel{03}{[+i]}\,\stackrel{12}{[+]}|\stackrel{56}{(+)}\,\stackrel{78}{(+)}|| 
\stackrel{9 \;10}{(+)}\;\;\stackrel{11\;12}{[+]}\;\;\stackrel{13\;14}{[+]}$ 
&$0$&$\frac{1}{2}$&$0$&$-\frac{1}{2}$&$-\frac{1}{2}$
  \\
$II$& 
%$u^{c1}_{R\,8}$&    $ \stackrel{03}{(+i)}\,\stackrel{12}{(+)}|\stackrel{56}{(+)}\,
%\stackrel{78}{(+)}||\stackrel{9 \;10}{(+)}\;\;\stackrel{11\;12}{(-)}\;\;
%\stackrel{13\;14}{(-)}$ & 
 $\nu_{R\,8}$&
 $ \stackrel{03}{(+i)}\,\stackrel{12}{(+)}|\stackrel{56}{(+)}\,\stackrel{78}{(+)}|| 
 \stackrel{9 \;10}{(+)}\;\;\stackrel{11\;12}{[+]}\;\;\stackrel{13\;14}{[+]}$ 
 &$0$&$\frac{1}{2}$&$0$&$\frac{1}{2}$&$-\frac{1}{3}$
 \\ 
 \hline 
 \end{tabular}}\label{Table III.} 
% \end{center}
\end{tiny}
\end{table}

All the families of Table~\ref{Table III.} and the family members of the eighth family in 
Table~\ref{Table so13+1.} are in the massless basis. 

The scalar fields, which are the gauge scalar fields  of  $\vec{\tilde{N}}_{R}$ and  
$\vec{\tilde{\tau}}^{2}$, couple only to the four families  which are doublets with respect to 
 these two groups. 
The scalar fields which are the gauge scalars  of  $\vec{\tilde{N}}_{L}$ and  
$\vec{\tilde{\tau}}^{1}$ 
couple only to the four families  which are doublets with respect to these last two groups. 

After the electroweak phase transition, caused by the scalar fields with the space index $(7,8)$%
~\cite{pikanorma,gn2015,JMP2015,IARD2016}, the two groups of four families become massive. 
The lowest of the two groups of four families contains the observed three, while the fourth family 
remains to be measured. The lowest of the upper four families is the 
candidate to form the dark matter~\cite{IARD2016,gn}.

\subsection{Vector gauge fields and scalar gauge fields in the {\it spin-charge-family} theory}
\label{vectorscalar}
%

% Vektorska polja in sklopitev spinorskimi in skalarnimi
%pred zlomitvijo simetrij,

In the {\it spin-charge-family} theory~\cite{IARD2016,norma2014MatterAntimatter,JMP2015},
like in all the Kaluza-Klein like theories, %~\cite{mil,zelenaknjiga}, 
either vielbeins or spin connections can be used to represent  the vector gauge fields in $d=(3+1)$ 
space, when space with $d\ge5$ has large enough symmetry and no strong spinor source is present. 
This is proven in Ref.~\citenum{EPJC2017} and the references therein. 
There are the superposition of $\omega_{stm}, m=(0,1,2,3), (s,t) \ge5$, which are used in the 
{\it spin-charge-family} theory to represent vector gauge fields --- $A^{Ai}_{m} (=\sum_{s,t} 
c^{Ai}{}_{st}\,\omega^{st}{}_{m}$) --- in $d=(3+1)$  in the low energy regime. Here $Ai$ 
represent the quantum numbers of the corresponding subgroups, expressed by the operators 
$S^{st}$ in Eqs.~(\ref{so42}, \ref{so64}).  Coefficients $c^{Ai}{}_{st}$  can be read from 
 Eqs.~(\ref{so42},\ref{so64}). These vector gauge fields manifest the properties of all the
directly and indirectly observed gauge fields%
~\footnote{In the {\it spin-charge-family} theory there 
are, besides the vector gauge fields of ($\vec{\tau}^{1}$,  $\vec{\tau}^{3}$), Eqs.~(\ref{so42},%
\ref{so64}), also the vector gauge fields of $\vec{\tau}^{2}$, 
Eq.~(\ref{so42}),  and $\tau^{4}$, Eq.~(\ref{so64}). The vector gauge fields of $\tau^{21}$,
$\tau^{22}$ and $Y'=\tau^{23} - \tan \theta_{2}\, \tau^4$ gain masses when interacting with the 
condensate~\cite{IARD2016} (and the references therein) at around $10^{16}$ GeV, while the
vector gauge field of the hyper charge $Y= \tau^{23} + \tau^4$ remains massless, together with
the gauge fields of $\vec{\tau}^{1}$ and $\vec{\tau}^{3}$, manifesting at low energies properties,
postulated by the {\it standard model}.}.

In the {\it spin-charge-family} theory also the scalar fields~\cite{norma2014MatterAntimatter,%
IARD2016,JMP2015,gmdn07,gn2015,EPJC2017} have the origin in the spin connection field, in
$\omega_{st s'} $ and $\tilde{\omega}_{st s'}, (s,t,s') \ge5$. These scalar fields offer the 
explanation for the Higgs's scalar and the Yukawa couplings of the {\it standard model}~\cite{gmdn07,%
IARD2016}. 

Both, scalar and vector gauge fields, follow from the simple starting action of the 
{\it spin-charge-family} presented in Eq.~(\ref{wholeaction}).

The Lagrange function for the vector gauge fields follows from the action for the curvature $R$ in 
Eq.~(\ref{wholeaction}) and manifests in the case of the flat $d=(3+1)$ space as assumed by
the {\it standard model}: $L_v= - \frac{1}{4}\, \sum_{\substack{A, i,m,n}}
 F^{Ai}{}_{mn} F^{A i \,mn}$, $F^{Ai}{}_{mn} = \partial_{m} A ^{Ai}_{n}- 
\partial_{n} A ^{Ai}_{m} - i f^{Aijk} \, A^{Aj}_{m} \,A^{Ak}_{n}$, with
\begin{eqnarray}
\label{vector}
A^{Ai}{}_{m}&=& \sum_{s, t} \,c^{Ai}{}_{s t}\, \omega^{s t}{}_{m}\,, \nonumber\\
\tau^{Ai} &=&  \sum_{s, t}\, c^{Ai s t} \,M_{s t}\,, \quad \quad 
M_{s t} = S_{st} + L_{st}\,.
\end{eqnarray}
In the low energy regime only $ S_{st}$ manifest. These expressions can be found in 
Ref.~\cite{EPJC2017}, Eq. (25), for example, and the references therein.

From Eq.~(\ref{wholeaction}) we read the interaction between fermions, presented in 
Table~\ref{Table so13+1.}, and the corresponding vector gauge fields in flat $d=(3+1)$ space.
\begin{eqnarray}
\label{fermionvector}
{\mathcal L}_{fv} &=&  \bar{\psi}\gamma^{m} (p_{m}- \sum_{A,i}\;\tau^{Ai} 
A^{Ai}_{m}) \psi \,.
\end{eqnarray}

Particular superposition of spin connection fields, either $\omega_{st s'}$ or $\tilde{\omega}_{ab s'}$,
 $(s,t,s') \ge 5, (a,b)=(0,\cdots,8)$, with the 
scalar space index $s'=(7,8)$, % with respect to $d=(3+1)$,
 manifest at low energies as the scalar fields, which contribute to the masses of  the family members. 
The superposition of the scalar fields $\omega_{st t"} $ with the space index $ t''=(9,\cdots,14)$ 
contribute to the transformation of matter into antimatter and back, causing in the presence of the 
condensate~\cite{norma2014MatterAntimatter,IARD2016} the matter-antimatter asymmetry  of
 our universe.
The interactions of all these scalar fields with fermions follow from Eq.~(\ref{wholeaction})
\begin{eqnarray}
\label{fermionscalar}
{\mathcal L}_{fs} &=& \{ \sum_{s=7,8}\;  \bar{\psi} \gamma^{s} p_{0s} \; \psi \} 
+ \nonumber\\ 
& & \{ \sum_{t=5,6,9,\dots, 14}\;  \bar{\psi} \gamma^{t} p_{0t} \; \psi \}\,, 
\end{eqnarray}
where $p_{0s} =  p_{s}  - \frac{1}{2}  S^{s' s"} \omega_{s' s" s} - 
                    \frac{1}{2}  \tilde{S}^{ab}   \tilde{\omega}_{ab s}$, 
$p_{0t}   =    p_{t}  - \frac{1}{2}  S^{t' t"} \omega_{t' t" t} - 
                    \frac{1}{2}  \tilde{S}^{ab}   \tilde{\omega}_{ab t}$,                    
with $ m \in (0,1,2,3)$, $s \in (7,8),\, (s',s") \in (5,6,7,8)$, $(a,b)$ (appearing in
 $\tilde{S}^{ab}$) run within  either $ (0,1,2,3)$ or $ (5,6,7,8)$, $t$ runs $ \in (5,\dots,14)$, 
$(t',t")$ run either $ \in  (5,6,7,8)$ or $\in (9,10,\dots,14)$. 
The spinor function $\psi$ represents all family members of all the $2^{\frac{7+1}{2}-1}=8$ 
families presented in Table~\ref{Table III.}.

There are the superposition of the scalar fields $\omega_{s' s" s}$ --- ($A^{Q}_{\pm}$,  
$A^{Q'}_{\pm}$,
$A^{Y'}_{\pm}$)~\footnote{$Q: =  \tau^{13} + Y$,  $Q':= -Y \tan^2\vartheta_1 +
\tau^{13}$, $Q':= - \tan^2\vartheta_1 Y + \tau^{13}$, $Y:= \tau^{4} + \tau^{23}$, 
$Y':= - \tan^2\vartheta_2 \tau^{4} + \tau^{23}$,
$Q: =  \tau^{13} + Y$, and correspondingly $A^{Q}_{s} = \sin \vartheta_{1} \,A^{13}_{s} +
 \cos \vartheta_{1} \,A^{Y}_{s}$, 
$A^{Q'}_{s}  = \cos \vartheta_{1} \,A^{13}_{s} - \sin \vartheta_{1} \,A^{Y}_{s}$,
$A^{Y'}_{s} = \cos \vartheta_{2} \,A^{23}_{s} - \sin \vartheta_{2} \,A^{4}_{s}$,
$A^{4}_{s}  = - (\omega_{9\,10\,s} + \omega_{11\,12\,s} + \omega_{13\,14\,s})$,
$A^{13}_{s} =  (\omega_{56 s}- \omega_{78 s})$, 
$A^{23}_{s}=(\omega_{56 s}+ \omega_{78 s})$,
with $(s\in (7,8))$~(Re.~\cite{JMP2015}, Eq. (A9)). } --- and the superposition of 
$\tilde{\omega}_{s' s'' s}$ ---
 ($\vec{\tilde{A}}^{\tilde{N}_{L}}_{\pm}$, $\vec{\tilde{A}}^{\tilde{1}}_{\pm}$,
 $\vec{\tilde{A}}^{\tilde{N}_{R}}_{\pm}$, $\vec{\tilde{A}}^{\tilde{2}}_{\pm}$)~\footnote{
$\vec{\tilde{A}}^{\tilde{1}}_{s} = 
(\tilde{\omega}_{\tilde{5} \tilde{8}s}-  \tilde{\omega}_{\tilde{6} \tilde{7}s},\, 
\tilde{\omega}_{\tilde{5} \tilde{7}s}+  \tilde{\omega}_{\tilde{6} \tilde{8}s}, \,
\tilde{\omega}_{\tilde{5} \tilde{6}s}-
  \tilde{\omega}_{\tilde{7} \tilde{8}s})$,  $\vec{\tilde{A}}^{\tilde{N}_{\tilde{L}}}_{s} =
 (\tilde{\omega}_{\tilde{2} \tilde{3}s}+i\,
  \tilde{\omega}_{\tilde{0} \tilde{1}s}, \,    \tilde{\omega}_{\tilde{3} \tilde{1}s}+i\,
  \tilde{\omega}_{\tilde{0} \tilde{2}s},  \,   \tilde{\omega}_{\tilde{1} \tilde{2}s}+i\,
  \tilde{\omega}_{\tilde{0} \tilde{3}s})$, $ \vec{\tilde{A}}^{\tilde{2}}_{s} = 
(\tilde{\omega}_{\tilde{5} \tilde{8}s}+  \tilde{\omega}_{\tilde{6} \tilde{7}s}, \,
\tilde{\omega}_{\tilde{5} \tilde{7}s}-   \tilde{\omega}_{\tilde{6} \tilde{8}s},\,
 \tilde{\omega}_{\tilde{5} \tilde{6}s}+  \tilde{\omega}_{\tilde{7} \tilde{8}s})\,$  and
$\vec{\tilde{A}}^{\tilde{N}_{\tilde{R}}}_{s} = (\tilde{\omega}_{\tilde{2} \tilde{3}s}-i\,
  \tilde{\omega}_{\tilde{0} \tilde{1}s}, \,    \tilde{\omega}_{\tilde{3} \tilde{1}s}-i\,
  \tilde{\omega}_{\tilde{0} \tilde{2}s}, \,    \tilde{\omega}_{\tilde{1} \tilde{2}s}-i\,
  \tilde{\omega}_{\tilde{0} \tilde{3}s})\,$, where $(s\in (7,8))$~(Ref.~\cite{JMP2015}, Eq. (A8)).}
% (Eq.~(\ref{Atildeomegas} 
--- which determine mass terms of family members of spinors after the electroweak break. 
I shall use $A^{Ai}_{\pm} $ to represent all the scalar fields, which determine masses of family 
members, the Yukawa couplings  and the masses of the weak boson vector fields, $A^{Ai}_{\pm} =$
($\sum_{A,i, a,b} c^{A i s t} (\omega_{s t 7} \pm i \omega_{s t 8}) $ as well as $= \sum_{A,i, a,b} 
c^{A i s t} (\tilde{\omega}_{a b 7} \pm i \omega_{a b 8})$.

The part of the first term of Eq.~(\ref{fermionscalar}), in which summation runs over 
the space index $s = ($7,8$)$ --- $\sum_{s=7,8}  \bar{\psi} 
\gamma^{s} p_{0s} \psi$ --- determines after the electroweak break masses of the two groups of 
four families. The highest of the lower four families is predicted to be observed at the 
L(arge)H(adron)C(ollider)~\cite{gn2015}, the lowest of the higher four families is explaining the origin 
of the dark matter~\cite{gn}. 

The scalar fields in the part of the second term of Eq.~(\ref{fermionscalar}), in which summation runs 
over the space index $t =$ ($9,\cdots, 14$) --- $\sum_{t=9,\cdots,14}  \bar{\psi} 
\gamma^{t} p_{0t} \psi$  --- cause transitions from anti-leptons into quarks and anti-quarks into quarks 
and back, transforming antimatter into matter and back. In the expanding universe the condensate 
of two right handed neutrinos breaks this matter-antimatter symmetry, explaining the matter-antimatter
asymmetry of our universe~\cite{norma2014MatterAntimatter}.

Spin connection fields $\omega_{st s'}$ and $\tilde{\omega}_{st s'}$ interact also with vector 
gauge fields and among themselves~\cite{EPJC2017}. These interactions can be red from 
Eq.~(\ref{wholeaction}). 

\section{Discussions and open problems}
\label{openproblems}

The {\it spin-charge-family}  theory is offering the next step beyond both {\it standard models}, 
by explaining:\\
  i.  The origin of charges of the (massless) family members and the relation between their  
charges and spins.  The theory, namely, starts in  $d=(13+1)$  with the simple action for spinors, which 
interact with the gravity only (Eq.\ref{wholeaction}) (through the vielbeins and the two kinds of the 
spin connection fields), while one fundamental representation of $SO(13,1)$ contains, if analyzed with
respect to the subgroups $SO(3,1), SU(3), SU(2)_I, SU(2)_{II}$ 
and $U(1)_{II}$ of the group $SO(13,1)$, all the quarks and anti-quarks and all the leptons and 
anti-leptons with the properties assumed by the {\it standard model}, relating handedness and charges 
of spinors as well as of anti-spinors (Table~\ref{Table so13+1.}).\\
  ii.  The origin of families of fermions, since spinors carry two kinds of spins (Eq.~(\ref{twoclifford})) --- 
the Dirac $\gamma^a$ and $\tilde{\gamma}^a$. In $d=(3+1)$ $\gamma^a$  take care  of the 
observed spins and charges,  $\tilde{\gamma}^a$ take care  of families~(Table~\ref{Table III.}).  \\
 iii.  The origin of the massless vector gauge fields of the observed charges, represented by the 
superposition of the spin connection fields 
$\omega_{st m}, (s,t)\ge 5, m\le3$~\cite{EPJC2017,IARD2016,JMP2015}. \\ 
 iv. The origin of masses of family members and of heavy bosons. The superposition of 
$\omega_{st s'}, (s,t)\ge 5, s'=(7,8)$ and the superposition of $\tilde{\omega}_{ab s'}, (a,b) =
(0,\cdots,8), s' =(7,8)$ namely gain at the 
electroweak break constant values, determining correspondingly masses of the spinors (fermions) and 
of the heavy bosons, explaining~\cite{IARD2016,JMP2015,gn2015} the origin of the Higgs's scalar and  
the Yukawa couplings of the {\it standard model}.\\ 
 v.  The origin of the matter-antimatter asymmetry~\cite{norma2014MatterAntimatter}, since the 
superposition of  $\omega_{st s'}, s'\ge9$, cause transitions from anti-leptons into quarks and 
anti-quarks into quarks and back, while the appearance of the scalar condensate in the expanding 
universe breaks the CP symmetry, enabling the existence of matter-antimatter asymmetry. \\
vi.  The origin of the dark matter, since there are two groups of decoupled four families in the 
low energy regime. The neutron  made of quarks of the stable of the upper four families explains 
the appearance of the dark matter~\cite{gn}~\footnote{We followed in Ref.~\cite{gn} freezing out 
of the fifth family quarks and anti-quarks in the expanding universe to see whether baryons of the fifth 
family quarks are the candidates for the dark matter.}.\\
vii. The origin of the triangle anomaly cancellation in the standard model. All the quarks and anti-quarks 
and leptons and anti-leptons, left and  right handed, appear  within one fundamental 
representation of $SO(13,1)$~\cite{IARD2016,JMP2015}.\\
viii. The origin of all the gauge fields. The {\it spin-charge-family}  theory unifies the gravity with all the
 vector and scalar gauge fields, since  
in the starting action there is only gravity~(Eq.~(\ref{wholeaction})), represented by the vielbeins and 
the two kinds of the spin connection fields, which in the low energy regime manifests in $d=(3+1)$ as the 
ordinary gravity and all the directly and indirectly observed vector and scalar gauge fields~\cite{EPJC2017}. 
If there is no spinor condensate present, only one of the three fields is the propagating field (both spin connections
are expressible with the vielbeins). In the presence of the spinor fields the two spin connection fields 
differ among themselves (Ref.~\citenum{EPJC2017}, Eq.~(4), and the references therein). 

The more work is done on the {\it spin-charge-family}  theory, the more answers to the open questions 
of both {\it standard models} is the theory offering. 

There are, of course, still open questions (mostly common to all the models) like:\\

\noindent
{\bf a.}  How has our universe really started?  The {\it spin-charge-family}  theory  assumes $d=(13 +1)$,
but how "has the universe decided" to start with $d=(13 +1)$? If starting at $d = \infty$, how 
can it come to $(13+1)$ with the massless Weyl representation of only one handedness? We have
studied in a toy model the break of symmetry from $d=(5+1)$ into $(3+1)$~\cite{DHN}, finding that 
there is the possibility that spinors of one handedness remain massless after this break. This study gives 
a hope that breaking the symmetry from $(d-1) +1$, where $d$ is even  and $\infty$, could go, if the 
jump of $(d-1) +1$ to $((d-4)-1) +1$ would be repeated as twice the break suggested in 
Ref.~\cite{DHN}. These jumps should then be repeated all the way from $d$ 
$=\infty$ to $d=(13 +1)$.\\ 
\noindent
{\bf b.}  What did "force" the expanding universe to break the symmetry of $SO(13,1)$ to $SO(7,1)$ 
$\times SU(3)\times U(1)_{II}$ and then further to $SO(3,1)$ $\times SU(2)\times SU(3) \times 
U(1)_{I}$ and finally to $SO(3,1)\times SU(3) \times U(1)$? \\
From phase transitions of ordinary matter we know that changes of temperature and 
pressure lead a particular matter into a phase transition, causing that constituents of the 
matter (nuclei and electrons) rearrange, changing the symmetry of space. \\
In expanding universe the temperature and pressure change, forcing spinors  to make condensates 
(like it is the condensate of the two right handed neutrinos in the {\it spin-charge-family}
theory~\cite{JMP2015, norma2014MatterAntimatter,IARD2016}, which gives masses to vector gauge 
fields of $SU(2)_{II}$, breaking $SU(2)_{II} \times U(1)_{II}$ into $U(1)_{I}$).  There might be also 
vector gauge fields causing a change of  the symmetry (like does the colour vector gauge fields, which 
"dress" quarks and anti-quarks and bind them to massive colourless baryons and mesons of the ordinary,  
mostly the first family, matter). 
Also scalar gauge fields might cause the break of the symmetry of the space (as this do the 
superposition  of $\omega_{s't' s}$  and the superposition of $\tilde{\omega}_{ab s}$, 
$s=(7,8), (s', t") \ge5, (a,b)=(0,\cdots,8)$ in the {\it spin-charge-family} theory~\cite{IARD2016,%
JMP2015} by gaining constant values in $d=(3+1)$ and breaking correspondingly also the symmetry of
 the coordinate space in $d\ge5$). \\
All these remain to be studied. \\
\noindent
{\bf c.} What is the scale of the electroweak phase transition? How higher is this scale in comparison 
with the colour phase transition scale? If the colour phase transition scale is at around $1$ GeV (since
the first family quarks contribute to baryons masses around $1$ GeV), is the electroweak scale at around 
$1$ TeV (of the order of the mass of Higgs's scalar) or this scale is much higher, possibly at the
unification scale (since the {\it spin-charge-family} theory predicts two decoupled groups of four families 
and several scalar fields --- twice two triplets and three singlets~\cite{JMP2015,gn2015,IARD2016})?\\

\noindent
{\bf d.} There are several more open questions. Among them are 
the origin of the dark energy,  the appearance of fermions, the origin of inflation of the universe, 
quantization of gravity, and several others. Can the {\it spin-charge-family} theory be --- while predicting 
the fourth family to the observed three, several scalar fields, the fifth family as the origin of the dark 
matter,  the scalar fields transforming anti-leptons into quarks and anti-quarks into quarks and back and 
the condensate which break this symmetry --- the first step, which
can hopefully show the way to next steps?
\appendix{Some useful formulas and relations are presented~\cite{IARD2016,pikanorma}}
\label{formulas}

\begin{eqnarray}
\label{stildestrans}
S^{ac}\stackrel{ab}{(k)}\stackrel{cd}{(k)} &= -\frac{i}{2} \eta^{aa} \eta^{cc} 
\stackrel{ab}{[-k]}\stackrel{cd}{[-k]}\,,\,\quad\quad
\tilde{S}^{ac}\stackrel{ab}{(k)}\stackrel{cd}{(k)} = \frac{i}{2} \eta^{aa} \eta^{cc} 
\stackrel{ab}{[k]}\stackrel{cd}{[k]}\,,\,\nonumber\\
S^{ac}\stackrel{ab}{[k]}\stackrel{cd}{[k]} &= \frac{i}{2}  
\stackrel{ab}{(-k)}\stackrel{cd}{(-k)}\,,\,\quad\quad
\tilde{S}^{ac}\stackrel{ab}{[k]}\stackrel{cd}{[k]} = -\frac{i}{2}  
\stackrel{ab}{(k)}\stackrel{cd}{(k)}\,,\,\nonumber\\
S^{ac}\stackrel{ab}{(k)}\stackrel{cd}{[k]}  &= -\frac{i}{2} \eta^{aa}  
\stackrel{ab}{[-k]}\stackrel{cd}{(-k)}\,,\,\quad\quad
\tilde{S}^{ac}\stackrel{ab}{(k)}\stackrel{cd}{[k]} = -\frac{i}{2} \eta^{aa}  
\stackrel{ab}{[k]}\stackrel{cd}{(k)}\,,\,\nonumber\\
S^{ac}\stackrel{ab}{[k]}\stackrel{cd}{(k)} &= \frac{i}{2} \eta^{cc}  
\stackrel{ab}{(-k)}\stackrel{cd}{[-k]}\,,\,\quad\quad
\tilde{S}^{ac}\stackrel{ab}{[k]}\stackrel{cd}{(k)} = \frac{i}{2} \eta^{cc}  
\stackrel{ab}{(k)}\stackrel{cd}{[k]}\,. 
\end{eqnarray}
\begin{eqnarray}
\stackrel{ab}{(k)}\stackrel{ab}{(k)}& =& 0\,,  \quad \stackrel{ab}{(k)}\stackrel{ab}{(-k)}
= \eta^{aa}  \stackrel{ab}{[k]}\,, \quad \stackrel{ab}{(-k)}\stackrel{ab}{(k)}=
\eta^{aa}   \stackrel{ab}{[-k]}\,,\;\;
\stackrel{ab}{(-k)} \stackrel{ab}{(-k)} = 0\,, \nonumber\\
\stackrel{ab}{[k]}\stackrel{ab}{[k]}& =& \stackrel{ab}{[k]}\,,  \;\;
\stackrel{ab}{[k]}\stackrel{ab}{[-k]}= 0\,, \;\;\quad \quad  \quad \stackrel{ab}{[-k]}\stackrel{ab}{[k]}=0\,,
 \;\;\quad \quad \quad \quad \stackrel{ab}{[-k]}\stackrel{ab}{[-k]} = \stackrel{ab}{[-k]}\,,
 \nonumber\\
\stackrel{ab}{(k)}\stackrel{ab}{[k]}& =& 0\,, \;\; \quad \stackrel{ab}{[k]}\stackrel{ab}{(k)}
=  \stackrel{ab}{(k)}\,, \quad \quad \quad \stackrel{ab}{(-k)}\stackrel{ab}{[k]}=
 \stackrel{ab}{(-k)}\,,\quad \quad \quad 
\stackrel{ab}{(-k)}\stackrel{ab}{[-k]} = 0\,,
\nonumber\\
\stackrel{ab}{(k)}\stackrel{ab}{[-k]}& =&  \stackrel{ab}{(k)}\,,
\quad \quad \stackrel{ab}{[k]}\stackrel{ab}{(-k)} =0,  \;\; 
\quad \stackrel{ab}{[-k]}\stackrel{ab}{(k)}= 0\,, \quad \quad \quad \quad
\stackrel{ab}{[-k]}\stackrel{ab}{(-k)} = \stackrel{ab}{(-k)}.
\label{graphbinoms}
\end{eqnarray}
\begin{eqnarray}
\label{plusminus}
N^{\pm}_{+}         &=& N^{1}_{+} \pm i \,N^{2}_{+} = 
 - \stackrel{03}{(\mp i)} \stackrel{12}{(\pm )}\,, \quad N^{\pm}_{-}= N^{1}_{-} \pm i\,N^{2}_{-} = 
  \stackrel{03}{(\pm i)} \stackrel{12}{(\pm )}\,,\nonumber\\
\tilde{N}^{\pm}_{+} &=& - \stackrel{03}{\tilde{(\mp i)}} \stackrel{12}{\tilde{(\pm )}}\,, \quad 
\tilde{N}^{\pm}_{-}= %\tilde{N}^{1}_{-} \pm i\,\tilde{N}^{2}_{-} = 
  \stackrel{03} {\tilde{(\pm i)}} \stackrel{12} {\tilde{(\pm )}}\,,\nonumber\\ 
\tau^{1\pm}         &=& (\mp)\, \stackrel{56}{(\pm )} \stackrel{78}{(\mp )} \,, \quad   
\tau^{2\mp}=            (\mp)\, \stackrel{56}{(\mp )} \stackrel{78}{(\mp )} \,,\nonumber\\ 
\tilde{\tau}^{1\pm} &=& (\mp)\, \stackrel{56}{\tilde{(\pm )}} \stackrel{78}{\tilde{(\mp )}}\,,\quad   
\tilde{\tau}^{2\mp}= (\mp)\, \stackrel{56}{\tilde{(\mp )}} \stackrel{78}{\tilde{(\mp )}}\,.
\end{eqnarray}
%

%\begin{verbatim}

\end{document}